\documentclass[11pt, a4paper]{article}

\usepackage{color}
\usepackage{graphicx}
\usepackage{amssymb}
\usepackage{amsmath}
\usepackage[english]{babel}
\usepackage[T1]{fontenc}
\usepackage[noadjust]{cite}
\usepackage{hyperref} 

\usepackage[utf8]{inputenc}
\usepackage{authblk}

\newcommand\eeq{\end{equation}}
\newcommand\beq{\begin{equation}}
\newcommand\eea{\end{eqnarray}}
\newcommand\bea{\begin{eqnarray}}

\begin{document}
\linespread{1.1}

\title{ \color{red} \bf Stationary configurations \\ of the Standard Model Higgs potential:\\
electroweak stability and rising inflection point}

\author[1]{ {\Large Giuseppe Iacobellis} \thanks{iacobellis@fe.infn.it}}
\author[1,2]{ {\Large Isabella Masina} \thanks{masina@fe.infn.it}}

\affil[1]{\small Dip. di Fisica e Scienze della Terra, Ferrara University and INFN, Ferrara, Italy }
\affil[2]{${\rm CP}^{3}$ - Origins \& DIAS, Southern Denmark University, Odense, Denmark}

\date{}

\maketitle

\begin{abstract}
We study the gauge-independent observables associated with two interesting stationary configurations of the Standard Model Higgs potential (extrapolated to high energy according to the present state of the art, namely the NNLO):  
i) the value of the top mass ensuring stability of the SM electroweak minimum, and ii) the value of the Higgs potential at a rising inflection point. We examine in detail and reappraise the experimental and theoretical uncertainties which plague their determination, finding that: 
i) stability of the SM is compatible with the present data at the $1.5\,\sigma$ level; 
ii) despite the large theoretical error plaguing the value of the Higgs potential at a rising inflection point, application of such configuration to models of primordial inflation displays a $3\,\sigma$ tension with the recent bounds on the tensor-to-scalar ratio of cosmological perturbations.
\end{abstract}

\linespread{1.2}

\vskip 1.cm
\section{Introduction}

The extrapolation of the Standard Model (SM) Higgs potential at high energy via the Renormalization Group Equation (RGE) is an interesting task \cite{Hung:1979dn, Cabibbo:1979ay}.  On one hand, the SM can be considered valid up to some energy scale only if the electroweak minimum is stable, or at least metastable. On the other hand, the shape of the Higgs potential at high energy could have some impact  on the dynamics of the early universe. 
Two stationary configurations of the SM Higgs potential turn out to be particularly relevant: i) the case of two degenerate vacua \cite{Froggatt:1995rt}, that is the condition for electroweak vacuum stability, and ii) the case of a rising inflection point at high energy \cite{MasinaHiggsmass}, close to the Planck scale. 

Our goal is to perform a detailed and updated study of the gauge-independent observables associated with such stationary configurations, building on the recent progresses made both in the theoretical and the experimental side. 
We extrapolate the SM Higgs potential up to high energy according to the present state of the art, 
namely the Next-to-Next-to-Leading-Order (NNLO), and study in particular:  
i) the value of the top mass ensuring stability of the SM electroweak minimum, and ii) the value of the Higgs potential at the rising inflection point. We examine in detail and reappraise in a critical way the experimental and theoretical uncertainties which plague their determination.

The inputs necessary to carry out the extrapolation are: the low energy values of the three gauge couplings, 
the top quark and the Higgs masses. 
After the discovery of the Higgs boson \cite{atlas, cms}, a lot of work has been done in the direction of refining the calculation of the RGE-improved Higgs effective potential at high energy. 
As for the matching with the experimental inputs at low energy, progresses were done in refs. \cite{Degrassi, Bezrukov:2012sa,  Buttazzo:2013uya,Bednyakov:2015sca}. 
A better understanding on how to extract gauge-invariant observables \cite{DiLuzio:2014bua, Espinosa:2015qea} from the (gauge-dependent) effective potential is worth to be mentioned.
Particularly relevant for the sake of the present analysis are the insights about the effective potential expansion in the case that the Higgs quartic coupling is small \cite{Andreassen:2014eha, Andreassen:2014gha}: these progresses 
justify the method used in previous analysis \cite{Degrassi, Bezrukov:2012sa,  Buttazzo:2013uya} to carry out the calculation of the effective potential at NNLO.

On the experimental side, the situation has also changed a bit in the last years. 
It is worth to mention the better knowledge of the Higgs mass, now measured with significant precision by ATLAS and CMS \cite{Aad:2015zhl}: $m_H=125.09 \pm 0.21 (stat)\pm 0.11 (syst)$ GeV. 
On the other hand, it was recognized that the error in the determination of the strong coupling constant has been previously underestimated by a factor of two, so that the present world average is $\alpha_s=0.1181 \pm 0.0013$ \cite{Agashe:2015}. Another open issue much debated in the literature is the determination of the top pole mass from the Monte Carlo (MC) mass \cite{Moch:2014lka, Nason:2016tiy, Corcella:2015kth}; the combined measurements of the Tevatron and LHC give, for the latter, 
$m_t^{MC}=173.34 \pm 0.76$ GeV  \cite{ATLAS:2014wva}.

Taking into account all these developments, we update the analysis of refs. \cite{Degrassi, Bezrukov:2012sa, Buttazzo:2013uya, Alekhin:2012py, Masina:2012tz, Bezrukov:2014ina, Bednyakov:2015sca}, reappraising in a critical way the experimental and theoretical uncertainties which plague the determination of the top mass value ensuring stability of the SM electroweak minimum. 
We find that stability of the SM is consistent with the present data at the $1.5\,\sigma$ level. Previous claims of a stronger tension at the $2\,\sigma$ level \cite{Degrassi, Bezrukov:2012sa,Buttazzo:2013uya} are, in our opinion, due to the previous underestimation of the experimental error on $\alpha_s$.

We also examine in a systematic way the theoretical and experimental uncertainties plaguing the calculation of the value of the Higgs potential at a rising inflection point. It turns out that, despite the largeness of these uncertainties, applications of such configuration to models of primordial inflation are disfavored  at $3\,\sigma$ by the recent bounds on the tensor-to-scalar ratio of cosmological perturbations. Claims of an even stronger tension \cite{Notari:2014noa, Ballesteros:2015iua} are, in our opinion, due to an underestimation of the theoretical errors involved in the calculation.

The paper is organized as follows. In sec.\,\ref{sec-calc} we present the details of the NNLO calculation, as the matching, running and effective potential expansion. We introduce the two stationary configurations in sec.\,\ref{sec-gauge}, discussing the gauge invariance of the observables related to them. Sec.\,\ref{sec-two} is devoted to the study of two degenerate vacua, while sec.\,\ref{sec-inf} to the inflection point configuration. Comments about the effect of higher-dimensional operators from an unknown gravitational sector are made in sec.\,\ref{sec-grav}. The conclusions are drawn in sec.\,\ref{sec-concl}.

\section{Calculation's state of the art} 
\label{sec-calc}

The normalization of the Higgs quartic coupling $\lambda$ is chosen in this paper so that the potential for the physical Higgs $\phi$ contained in the Higgs doublet $\mathcal{H}=(0\quad(\phi+v)/\sqrt{2})^T$ is given, at tree-level, by
\begin{equation}
V(\phi)=\frac{\lambda}{6} \left(\left|\mathcal{H}\right|^2 - \frac{v^2}{2} \right)^2 \approx  \frac{\lambda}{24}  \phi^4\,,
\label{eq-Vtree}
\end{equation}
where $v=1/(\sqrt2 G_\mu)^{1/2}=246.221{\rm ~GeV}$ and $G_\mu$ is the Fermi constant from muon decay \,\cite{Agashe:2014kda}. 
The approximation in eq.\,(\ref{eq-Vtree}) holds when considering large field values. 
According to our normalization, the physical Higgs mass satisfies the tree-level relation $m_H^2= \lambda v^2 /3$.
In addition, the mass of the fermion $f$ reads, at tree-level,  $ m_f =   h_f  v/\sqrt{2}$, where $h_f$ denotes the associated Yukawa coupling.

In order to extrapolate the behavior of the Higgs potential at very high energies, we adopt the $\overline{\rm MS}$ scheme 
and consider the matching and RGE evolution for the relevant couplings which, in addition to the Higgs quartic coupling $\lambda$, are: the gauge couplings $g$, $g'$, $g_3$, the top Yukawa coupling, $h_t$, and the anomalous dimension of the Higgs field, $\gamma$. We then compute the   
RGE-improved Higgs effective potential.

As anticipated, we perform this calculation according to the present state of the art, namely the NNLO. Before discussing in detail the procedure associated to matching, running, and effective potential expansion, we review the basic ideas of the RGE to introduce our notation.

In applications where the effective potential $V_{\text{eff}}(\phi)$ at large $\phi$ is needed, as is the case for our analysis, 
potentially large logarithms appears, of the type $\log(\phi/\mu)$ where $\mu$ is the renormalization scale, which may spoil the applicability of perturbation theory.  The standard way to treat such logarithms is by means of the RGE.
The fact that, for fixed values of the bare parameters, the effective potential must be independent of the renormalization scale $\mu$, means that \cite{Coleman}
\beq
\left( \mu \frac{\partial}{\partial \mu} + \beta_i \frac{\partial}{\partial \lambda_i} - \gamma \frac{\partial}{\partial \phi}  \right) V_{\text{eff}}=0 \, ,
\eeq
where 
\beq
\beta_i = \mu \frac{d \lambda_i}{d \mu} \,\, , \,\, \gamma = -\frac{\mu}{\phi} \frac{d\phi}{d\mu}\,,
\eeq
are the $\beta$-functions corresponding to each of the SM couplings $\lambda_i$, and the anomalous dimension of the background field respectively.

The formal solution of the RGE is
\beq
V_{\text{eff}}(\mu,\lambda_i,\phi)= V_{\text{eff}}(\mu(t),\lambda_i(t),\phi(t))\,,
\eeq
where 
\beq
\mu(t)=e^t \mu  \, ,\,\, \phi(t) =e^{\Gamma(t)} \phi \,,\,\, \Gamma(t)=- \int_0^t \gamma(\lambda(t')) dt' \, ,
\label{eq-mu}
\eeq
and $\lambda_i(t)$ are the SM running couplings, determined by the equation
\beq
\frac{d \lambda_i(t)}{dt} = \beta_i (\lambda_i(t)) \, ,
\eeq
and subject to the boundary conditions $\lambda_i(0)=\lambda_i$.
The usefulness of the RGE is that $t$ can be chosen in such a way that the convergence of
perturbation theory is improved, which is the case for instance when $\phi(t)/\mu(t) = {\cal{O}} (1)$.

Since in our calculation the boundary conditions are given at the top quark mass, $m_t$, we will take $\mu=m_t$ in eq. (\ref{eq-mu}) from now on.

\subsection{Matching}
\label{sec-match}

In order to derive the values of the relevant parameters ($g$, $g'$, $g_3$, $h_t$, $\lambda$) at the top pole mass, $m_t$, we exploit the results of the most recent analysis about the matching procedure, performed by Bednyakov et al. \cite{Bednyakov:2015sca}.   
This paper uses as input parameters at $m_Z$ those from the 2014 release of the Particle Data Group (PDG) \cite{Agashe:2014kda}. 
Such parameters are evolved in the context of the SM as an effective theory with five flavors, and then matched to the six flavors theory at $m_t$. 
The procedure is carried out by including corrections up to ${\cal{O}}(\alpha_{em}^2)$, ${\cal{O}}(\alpha_{em} \alpha_s)$, ${\cal{O}}(\alpha_s^4)$, where $\alpha_{em}$ and  $\alpha_s$ are the fine structure constant and the strong gauge coupling respectively. 
The theoretical uncertainties in the results due to unknown higher-order corrections are estimated considering both scale variations and truncation errors. The matching is thus carried out at the NNLO (actually even slightly beyond for the strong gauge coupling contribution).

Although in our analysis we use the complete results of \cite{Bednyakov:2015sca} (see in particular their eq. (6) and table I), we provide here some simplified expressions which capture the dominant dependences and sources of uncertainty. It is well known that, for the sake of the present calculation, the most significant uncertainties are those associated with the determination of $g_3(m_t), h_t(m_t), \lambda(m_t)$, while uncertainties in the matching of $g(m_t), g'(m_t)$ are negligible. 
Let us consider the former three couplings in turn, following closely Bednyakov et al. \cite{Bednyakov:2015sca}, but updating it  when necessary by means of the latest (September 2015) release of the PDG \cite{Agashe:2015}. 

\begin{itemize}

\item 

The uncertainty in the value of $g_3(m_t)$ is essentially dominated by the experimental error on the value of 
$\alpha_s^{(5)}$, the strong coupling constant at $m_Z$ in the SM with five flavors: 
\beq
g_3(m_t) \simeq 1.1636 + 5.8 \times 10^{-3} \,\, \frac{\alpha_s^{(5)} - \alpha_s^{(5,exp)}}{\Delta \alpha_s^{(5,exp)}} \, ,
\eeq
where the present (PDG revised version of September 2015) \cite{Agashe:2015} world average experimental value, $\alpha_s^{(5,exp)}=0.1181$, and its associated $1\,\sigma$ error, 
$\Delta \alpha_s^{(5,exp)}=0.0013$,  have been used as reference values.
Notice that ref. \cite{Bednyakov:2015sca} used instead as reference value $\alpha_s^{(5,exp)}=0.1185$, with $1\,\sigma$ error given by $\Delta \alpha_s^{(5,exp)}=0.0006$  \cite{Agashe:2014kda};
previous analyses like e.g. \cite{Degrassi, Bezrukov:2012sa, Buttazzo:2013uya}, used $\alpha_s^{(5,exp)}=0.1184$, with $1\,\sigma$ error given by $\Delta \alpha_s^{(5,exp)}=0.0007$  \cite{PDG2012}. 
The experimental error is approximately doubled at present because of a previous underestimation of the uncertainty in the lattice results.

\item

The uncertainty on $\lambda(m_t)$ is dominated by the experimental uncertainty on the Higgs mass $m_H$ and by the theoretical uncertainty associated to the matching procedure (scale variation and truncation, here added in quadrature) 
\beq
\lambda(m_t)  \simeq 0.7554 +2.9 \times 10^{-3} \,\frac{m_H- m_H^{exp}}{\Delta m_H^{exp}} 
 \pm 4.8 \times 10^{-3}\,,
 \label{eqlambda}
\eeq
where we used as reference values the most recent ATLAS and CMS combination\footnote{Again, we update the result of \cite{Bednyakov:2015sca} by using the most recent LHC data, instead of $m_H^{exp}=125.7$ GeV and $\Delta m_H^{exp}=0.4$ GeV, quoted in the 2014 version of the PDG \cite{Agashe:2014kda}. Notice also that within our convention the value of $\lambda$ is six times the one of \cite{Bednyakov:2015sca}.}, $m_H^{exp}=125.09$ GeV, with $1\,\sigma$ error given by $\Delta m_H^{exp}=0.24$ GeV \cite{Aad:2015zhl}.
Notice that in eq.\,(\ref{eqlambda}) the theoretical error is pretty large, being equivalent to a $1.6\,\sigma$ variation in $m_H$. 
In the previous literature there is some difference about the size of the theoretical error:
for instance, the theoretical error of the recent analysis by Bednyakov et al. \cite{Bednyakov:2015sca} is about the half of the one quoted in the well-known analysis by Degrassi et al. \cite{Degrassi}, due to the inclusion \cite{Bednyakov:2015sca} of all corrections up to ${\cal{O}}(\alpha_{em}^2)$, ${\cal{O}}(\alpha_{em} \alpha_s)$, ${\cal{O}}(\alpha_s^4)$. On the other hand, Buttazzo et al. \cite{Buttazzo:2013uya} quote an error which is five times smaller than the one of ref. \cite{Degrassi}; according to our previous analysis \cite{Masina:2012tz}, the upper error is as small as the one of ref. \cite{Buttazzo:2013uya}, but the lower error is indeed consistent with ref. \cite{Degrassi}. Clearly, it would be worth to assess and further refine the present error in the matching of $\lambda$. 

\item 

The uncertainty on $h_t(m_t)$  is mainly affected by the experimental error on the top pole mass $m_t$ itself, but also the theoretical uncertainty 
associated to the matching procedure (scale variation and truncation, here added in quadrature) is sizable
\beq
h_t(m_t) \simeq 0.9359 +4.4 \times 10^{-3} \, \frac{m_t- m_t^{exp}}{\Delta m_t^{exp}}\pm 1.4 \times 10^{-3}\,,
\label{eqht}
\eeq
where we used as reference values those 
of the first joint Tevatron and LHC analysis, 
$m_t^{exp}=173.34$ GeV and $\Delta m_t^{exp}=0.76$ GeV \cite{ATLAS:2014wva}.
The theoretical uncertainty in eq.\,(\ref{eqht}) is consistent with the one of refs. \cite{Degrassi, Buttazzo:2013uya}\footnote{The latter quotes an error on the matching of $h_t$ which is three times smaller than the one in eq.\,(\ref{eqht}), but includes an error of ${\cal O}(\Lambda_{\rm QCD})$, i.e. about $0.3$ GeV, in the definition of $m_t$, to account for
non-perturbative uncertainties associated with the relation between the measured value of the top mass and the actual definition of the top pole mass.}.
Since the top mass is extracted by fitting MC computed distributions
to experimental data, what is really measured is a MC parameter, $m_t^{MC}$.
\end{itemize}

According to some authors \cite{Moch:2014tta}, although it is common to identify the top quark pole mass $m_t$ with $m_t^{MC}$, the uncertainty in the translation from the MC mass definition to a theoretically well-defined short distance mass definition at a low scale should currently be estimated to be of the order of $1$ GeV  \cite{Hoang:2008xm}; if this were the case \cite{Alekhin:2012py, Masina:2012tz,Bednyakov:2015sca}, the customary confidence ellipses with respect to $m_t^{MC}$ and $m_H$  (more on this in sec.\,\ref{sec-two}) should be taken with a grain of salt. 

Other authors \cite{Nason:2016tiy, Corcella:2015kth} argued that measurements relying on the reconstruction of top-decay products yield results which are actually close to the top quark pole mass, 
although there are theoretical uncertainties due to the approximations contained in the MC shower models,
namely missing loop and width corrections and colour-reconnection effects. 
The discrepancy between MC and pole masses was estimated in \cite{Fleming:2007xt}, by 
identifying the MC mass as a SCET jet mass, evaluated at a scale given by the shower infrared cutoff, 
i.e. ${\cal O}(1\, {\rm GeV})$, in $e^+ e^- \rightarrow t\bar t$ collisions. As discussed in \cite{Corcella:2015kth}, such a discrepancy amounts to about $200$ MeV.
In addition, the renormalon ambiguity affecting the pole mass, was recently estimated \cite{Nason:2016tiy} as the size of the last converging term in the $\overline{\rm MS}$/pole relation, obtained extrapolating to higher orders the 4-loop computation in \cite{Marquard:2015qpa}, and amounting to less than $100$ MeV.
The 4-loop correction was also obtained in semi-analytical form in ref. \cite{Kataev:2015gvt}, finding agreement with ref. \cite{Marquard:2015qpa}.

\subsection{Running}
\label{sec-run}

The $\beta$-functions can be organized as a sum of contributions with increasing number of loops:
\beq
 \frac{d}{d t} \lambda_i(t)=\kappa \beta_{\lambda_i}^{(1)}+\kappa^2  \beta_{\lambda_i}^{(2)} +\kappa^3  \beta_{\lambda_i}^{(3)}  + ...\, ,  \label{eq-RGE}
\eeq
where $\kappa = 1/(16 \pi^2)$ and the apex on the $\beta$-functions represents the loop order.

Here, we are interested in the RGE dependence of the couplings ($g$, $g'$, $g_3$, $h_t$, $\lambda$, $\gamma$). 
The 1-loop and 2-loop expressions for the $\beta$-functions in the SM are well known and can be found {\it e.g.} in Ford et al. \cite{Ford:1992}. 
The complete 3-loop $\beta$-functions for the SM have been computed more recently:
as for the SM gauge couplings, they have been presented by Mihaila, Salomon and Steinhauser  in refs. \cite{Mihaila:2012,Mihaila1};
as for $\lambda$, $h_t$ and the Higgs anomalous dimension, they have been presented by Chetyrkin and Zoller in refs. \cite{ChetyrkinZoller,Chetyrkin:2013}
and by Bednyakov et al. in refs. \cite{BednyakovPikelnerVelizhanin,BednyakovPikelnerVelizhanin1,Bednyakov:2013, Bednyakov:2014}.
The dominant 4-loop contribution to the running of the strong gauge coupling has been also computed recently, see refs. \cite{Zoller:2015tha,Bednyakov:2015}. In our NNLO analysis, we include all these contributions\footnote{We do not include the 4-loop contribution to the Higgs self-coupling, anomalous dimension of the Higgs field and top Yukawa $\beta$-functions calculated very recently by Chetyrkin and Zoller \cite{Chetyrkin:2016ruf}. A numerical estimate of these terms by the same authors leads to the conclusion that they are negligible with respect to the other sources of uncertainty.}.

\subsection{RGE-improved effective potential}\label{eff}

As already stressed, $t$ can be chosen in such a way that the convergence of
perturbation theory is improved. 
Without sticking, for the time being, to any specific choice of scale, the RGE-improved effective potential at high field values can be rewritten as
\beq
V_{\text{eff}}(\phi,t) \approx \frac{\lambda_{\text{eff}}(\phi,t)}{24} \phi^4\,,  
\eeq
where $\lambda_{\text{eff}} (\phi,t)$ takes into account wave function normalization and can be expanded as sum of tree-level plus increasing loop contributions: 
\beq
\lambda_{\text{eff}}(\phi,t)= e^{4 \Gamma(t)} 
\left[   \lambda(t) +  \lambda^{(1)}(\phi,t) +  \lambda^{(2)}(\phi,t) +... \right]\,.
\eeq

In particular, the 1-loop Coleman-Weinberg contribution \cite{Coleman:1973jx} is
\beq
\lambda^{(1)}(\phi,t)= 6 \frac{1}{(4 \pi)^2} \sum_p N_p \kappa^2_p(t) \left(  \log \frac{\kappa_p(t) e^{2 \Gamma(t)} \phi^2}{\mu(t)^2} -C_p \right)\,,    
\label{eqla1}
\eeq
where, generically, $p$ runs over the top quark, $W$, $Z$, Higgs and Goldstone bosons contributions.
The coefficients $N_p$, $C_p$, $\kappa_p$ are listed in the table below for the Landau gauge (see {\it e.g.} table 2 of ref. \cite{DiLuzio:2014bua} for a general $R_\xi$ gauge). 

\begin{table}[h!]
\vskip .5cm 
\centering
\label{my-label}
\begin{tabular}{  r | c  c  c   c  c      }
 $p$ & $t$ & $W$ & $Z$ & $\phi$  & $\chi$  \\
\hline
$N_p$         & $-12$ & $6$ & $3$  & $1$ & $3$   \\
$C_p$         &  $3/2$ & $5/6$  & $5/6$ & $3/2$ & $3/2$  \\
$\kappa_p$ &  $h^2/2$ & $g^2/4$ & $(g^2+g'^2)/4$ & $3\lambda$ & $\lambda$
\end{tabular}\caption{Coefficients for eq.\,(\ref{eqla1}) in the Landau gauge. }
\vskip .5cm 
\end{table}

The 2-loop contribution $\lambda^{(2)}(\phi, t)$ was derived by Ford et al. in ref. \cite{Ford:1992} and, in the limit $\lambda\rightarrow0$,  was cast in a more compact form in refs. \cite{Degrassi,Buttazzo:2013uya}. 
We verified, consistently with these works, that the error committed in this approximation is less than $10\%$ and can thus be neglected.

It is clear that when $\lambda(t)$ becomes negative, the Higgs and Goldstone contributions in eq.\,(\ref{eqla1}) are small but complex, 
and this represents a problem in the numerical analysis of the stability of the electroweak vacuum. Indeed, 
in refs. \cite{Degrassi,Buttazzo:2013uya} the potential was calculated at the 2-loop level, but setting to zero the Higgs and Goldstone contributions in eq.\,(\ref{eqla1}). In ref. \cite{Masina:2012tz} we rather decided to calculate the potential only at the tree-level because, for the sake of the analysis of the electroweak vacuum stability, the numerical difference with respect to the previous method is negligible.

Some authors \cite{Elias-Miro:2014pca, Andreassen:2014gha} recently showed that the procedure of refs. \cite{Degrassi,Buttazzo:2013uya} is actually theoretically justified when $\lambda$ is small (say $\lambda \sim \hbar$): in this case, the sum over $p$ does not have to include the Higgs and Goldstone's contributions, which rather have to be accounted for in the 2-loop effective potential,
which practically coincides with the expression derived in refs. \cite{Degrassi,Buttazzo:2013uya}. 

For the stationary configurations we are interested in -- two degenerate minima and a rising point of inflection -- it happens that $\lambda$ is small (and could be negative): we thus adopt the procedure outlined in \cite{Andreassen:2014gha}.

\section{On the gauge (in)dependence}
\label{sec-gauge}

Another aspect of the present calculation is represented by the well-known fact that the RGE-improved effective potential is gauge dependent, although the value of the top mass corresponding to the stability bound is not \cite{Buttazzo:2013uya}.
To introduce the issue, let us consider the argument presented by Di Luzio et al. in ref. \cite{DiLuzio:2014bua} for the case of two degenerate vacua, generalizing it to any stationary configuration (here we use $m_t$ instead of $m_H$). 

\begin{figure}[h!]
\vskip .5cm 
 \begin{center}
\includegraphics[width=12cm]{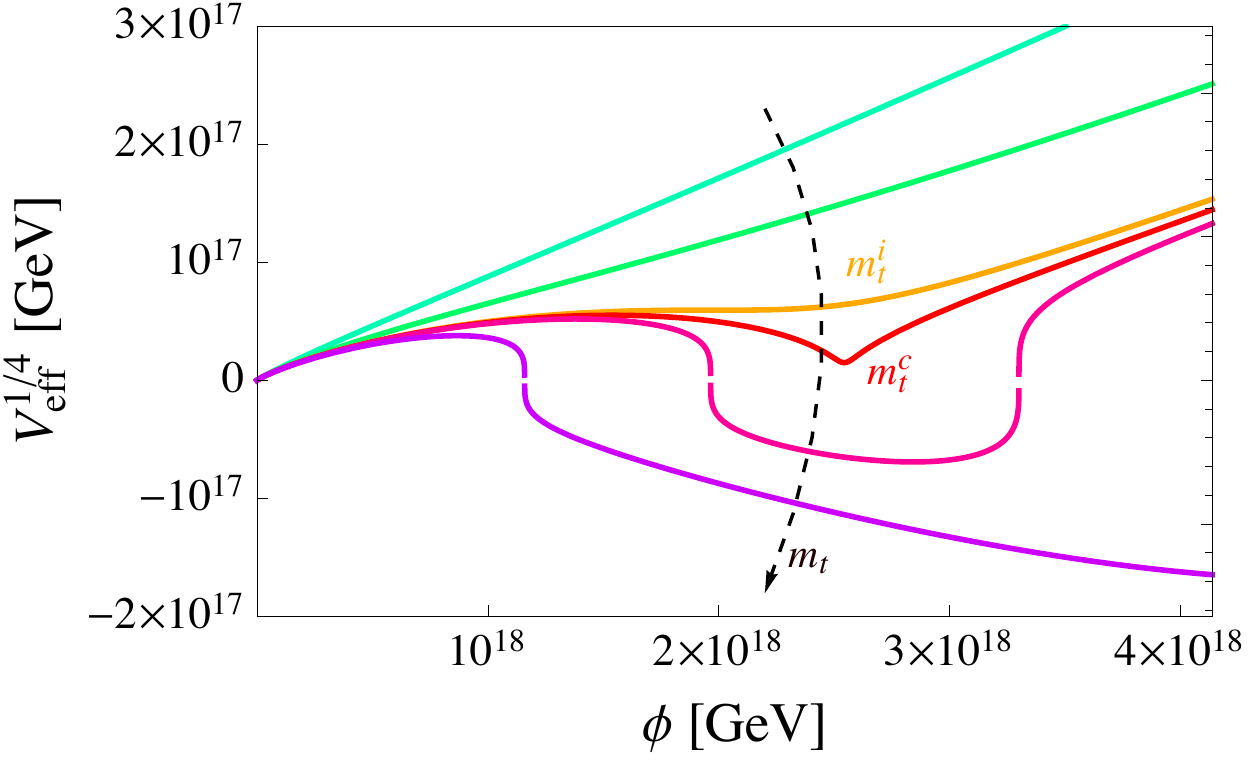} \end{center}
\caption{\baselineskip=15 pt  \small
Sketch of the  shape of SM effective potential at high energy in the Landau gauge, for increasing values of the top mass from top to bottom. Two peculiar stationary configuration can be identified: two degenerate vacua and a rising point of inflection, associated respectively to $m_t^c$ and $m_t^i$. 
 }
\label{fig-multi}
\vskip .5 cm
\end{figure}

Let us assume that all the parameters of the SM are exactly determined, except for the top mass.
After choosing the renormalization scale $t$, the RGE-improved effective potential,
$V_{\text{eff}}( \phi, m_t; \xi)$, is a function of $\phi$, the top mass $m_t$, and the gauge-fixing parameters, collectively 
denoted by $\xi$. One can think of $m_t$ as a free parameter, whose variation modifies the shape of the effective potential,
as sketched in fig.\,\ref{fig-multi} for the Landau gauge. Starting from top to bottom, the shape of the Higgs potential 
changes, going from stability to instability by increasing the top mass. 

The absolute stability bound on the top mass can be obtained by defining a critical mass, $m_t^c$, 
for which the value of the effective potential at the electroweak minimum, $\phi_{ew}$, and at
a second minimum,  $\phi_c > \phi_{ew}$, are the same:
\beq
\left.\frac{\partial V_{\text{eff}}}{\partial \phi}\right|_{\phi_{ew},m^c_t}  =\left.\frac{\partial V_{\text{eff}}}{\partial \phi}\right|_{\phi_c,m^c_t}  =0 \, , \,\,\, 
V_{\text{eff}} (\phi_{ew}, m^c_t;\xi) -V_{\text{eff}} ( \phi_c, m^c_t;\xi) =0\,.
\label{eq-c}
\eeq
Slightly reducing $m_t$, one finds another particular value of the top mass, $m^i_t$, such that the Higgs potential displays a rising point of inflection at $\phi_i > \phi_{ew}$:
\beq
\left.\frac{\partial V_{\text{eff}}}{\partial \phi}\right|_{\phi_{ew},m^i_t}  =\left.\frac{\partial V_{\text{eff}}}{\partial \phi}\right|_{\phi_i,m^i_t}  =0\,, \,\,\,  
\left.\frac{\partial^2 V_{\text{eff}}}{\partial \phi^2}\right|_{\phi_i,m^i_t}  =0\,.
\label{eq-ip}
\eeq

Due to the explicit presence of $\xi$ in the vacuum stability and/or inflection point conditions, 
it is not obvious a priori which are the physical (gauge-independent) observables entering the vacuum
stability and/or inflection point analysis. 

The basic tool, in order to capture the gauge-invariant content of the effective
potential is given by the Nielsen identity \cite{Nielsen:1975fs}
\beq
\left(\xi   \frac{\partial }{\partial \xi}  + C(\phi,\xi)   \frac{\partial  }{\partial \phi}  \right)V_{\text{eff}}(\phi,\xi)=0\,,
\eeq
where $C(\phi, \xi)$ is a correlator whose
explicit expression will not be needed for our argument.
The equation means that $V_{\text{eff}}(\phi,\xi)$ is constant along the characteristics of the equation,
which are the curves in the $(\phi, \xi)$ plane for which $d\xi= \xi /C(\phi,\xi) d\phi $.

In particular, the identity says that the effective potential is gauge independent where it is stationary.
Due to the fact that the value of the effective potential at any stationary point $\phi_s$ (as is the case for $\phi_{ew}$, $\phi_c$,  $\phi_i$) is gauge invariant
\beq
\left.\frac{\partial V_{\text{eff}}(\phi,\xi)}{\partial \phi}\right|_{\phi_s,m_t}  =0 \rightarrow
\left.\frac{\partial V_{\text{eff}}(\phi,\xi)}{\partial \xi}\right|_{\phi_s,m_t}  =0\,, 
\eeq
its value at the extremum can be calculated working in any particular gauge, say the Landau gauge
\beq
V_{\text{eff}}(\phi_s, m_t;\xi) = V_{\text{eff}}(\phi_s^L, m_t ; 0)\,,
\label{eq-s}
\eeq
where $\phi_{s}^{L}$ is the field evaluated in a stationary point in Landau gauge.

We want to check explicitly that, at the contrary of $\phi_s$, the particular values of $m_t$ ensuring criticality, $m^c_t$, and an inflection point, $m^i_t$, are gauge independent. Let us call them collectively $m^s_t$ ($s=c,i$), and denote by $\bar V_s$ the associated value of the effective potential in the Landau gauge:
\beq
V_{\text{eff}}(\phi_s, m^s_t;\xi) = V_{\text{eff}}(\phi_s^L, m^s_t ; 0) \equiv \bar V_s\,.
\label{eq-V}
\eeq
Inverting eq.\,(\ref{eq-V}) (together with the stationary condition) would yield gauge-dependent field 
and top mass values: $\phi_s=\phi_s (\xi)$ and $m^s_t=m^s_t(\xi)$.
We apply a total derivative with respect to $\xi$ to eq.\,(\ref{eq-s}) and obtain
\beq
\left.\frac{\partial V_{\text{eff}}}{\partial \xi}\right|_{\phi_s,m^s_t}  
+ \left.\frac{\partial V_{\text{eff}}}{\partial m_t}\right|_{\phi_s,m^s_t}   \frac{\partial m^s_t}{\partial \xi}
 + \left.\frac{\partial V_{\text{eff}}}{\partial \phi}\right|_{\phi_s,m^s_t}  \frac{\partial \phi_s}{\partial \xi} 
=0 \, ,
\label{eq-dem}
\eeq
where the third and first terms in the l.h.s. vanish because of the stationary condition and the Nielsen identity respectively.
Since in general $\left.\frac{\partial V_{\text{eff}}}{\partial m_t}\right|_{\phi_s,m^s_t}  \neq 0$, we obtain that $\frac{\partial m^s_t}{\partial \xi}=0$.
We can conclude that the peculiar values of $m_t$ ensuring stationary configurations, like two degenerate vacua or an inflection point, are gauge independent.

Notice that the above argument can be easily generalized to the case in which we treat as free parameters not only the top mass, but all other input parameters entering in the calculation of the effective potential, as for instance the Higgs mass and $\alpha_s$. Let us call them $\vec f=(m_t, m_H, \alpha_s, ...)$, so that $V_{\text{eff}}( \phi, \vec f; \xi)$. In this case, the generalization of eq. (\ref{eq-dem}) is simply:
\beq
\left.\frac{\partial V_{\text{eff}}}{\partial \xi}\right|_{\phi_s, \vec f^s}  
+\sum_i  \left.\frac{\partial V_{\text{eff}}}{\partial f_i}\right|_{\phi_s, \vec f^s}   \frac{\partial f^s_i}{\partial \xi}
 + \left.\frac{\partial V_{\text{eff}}}{\partial \phi}\right|_{\phi_s, \vec f^s}  \frac{\partial \phi_s}{\partial \xi} 
=0 \, .
\eeq
As before, the last and the first terms in the l.h.s. vanish because of the stationary condition and the Nielsen identity respectively.
Since in general $\left.\frac{\partial V_{\text{eff}}}{\partial f_i}\right|_{\phi_s, \vec f^s}  \neq 0$, we obtain $\frac{\partial f^s_i}{\partial \xi}=0$.
The peculiar values of the input parameters ensuring stationary configurations, like two degenerate vacua or an inflection point, are thus gauge independent.

Working in the Landau gauge is thus perfectly consistent in order to calculate the value  of the effective potential at a stationary point, 
$\bar V_s$, or the value of the top mass providing it\footnote{Of course, this applies also to the other input parameters, but here we focus on $m_t$ as it is the one associated to the largest uncertainty.}, $m^s_t$.
Nevertheless, one has to be aware that the truncation of the effective potential loop expansion at some loop order, introduces an unavoidable theoretical error both in $\bar V_s$ and in $m^s_t$. 
For this sake, it is useful to define the parameter $\alpha$ via
\beq
\mu(t)= \alpha \, \phi\, ,
\label{eq-alfa}
\eeq
and study the dependence of $\bar V_s$ and $m^s_t$ on $\alpha$. The higher the order of the loop expansion to be considered, the less the dependence on $\alpha$. This will be shown explicitly in the next two sections, where we study in detail the case of two degenerate vacua and the case of a rising inflection point at high energies, respectively.

\section{Two degenerate vacua}
\label{sec-two}

As discussed in the previous section, once $m_H$ and $\alpha_s^{(5)}$ have been fixed, 
the value of the top mass for which the SM displays two degenerate vacua, $m^c_t$, is a gauge invariant quantity.
This value is however plagued by experimental and theoretical errors, which we now discuss in turn.

\begin{figure}[t!]
\vskip 1.5cm 
 \begin{center}
\includegraphics[width=12cm]{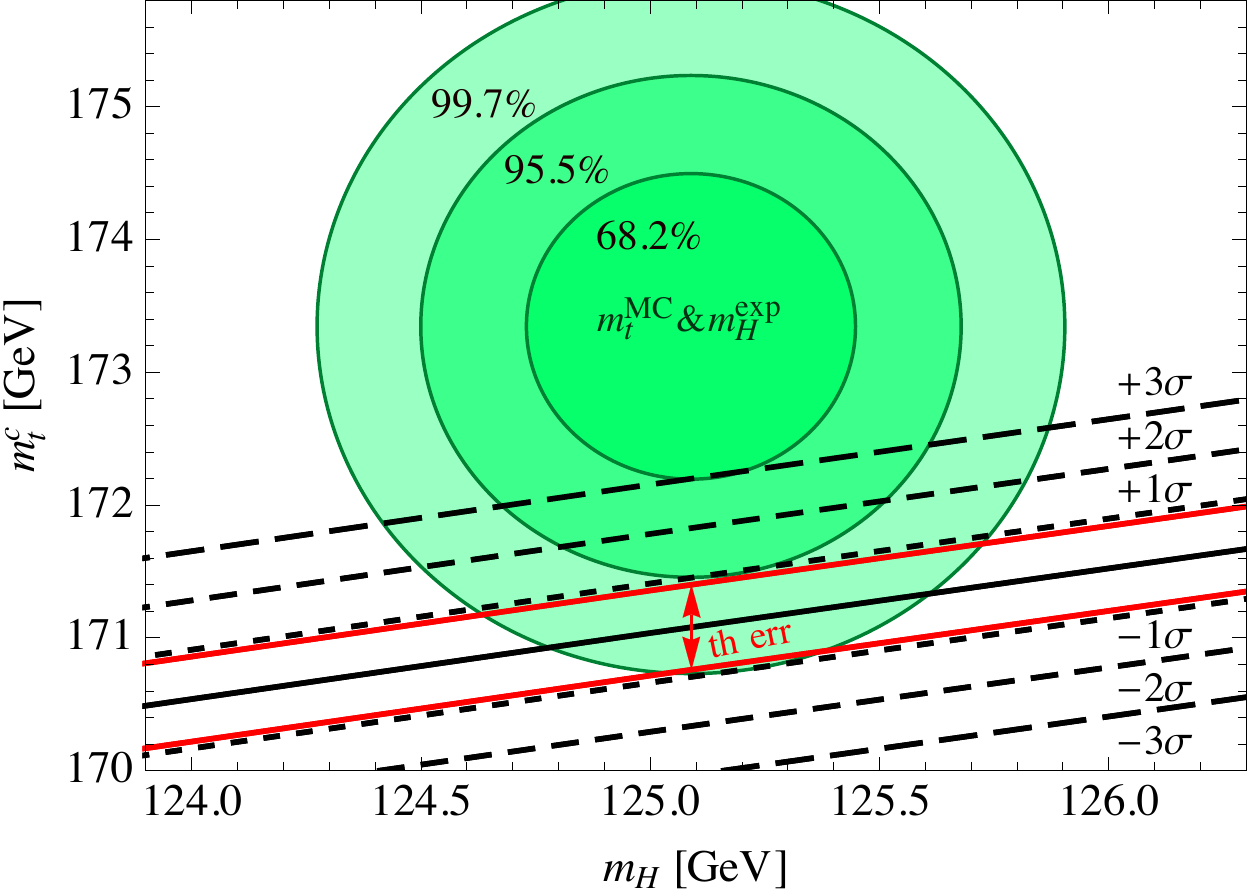}   
 \end{center}
\caption{\baselineskip=15 pt \small  
Lines for which the Higgs potential develops a second degenerate minimum at high energy. The solid line corresponds to the central value of $\alpha_s^{(5)}$; the dashed lines are obtained by varying $\alpha_s^{(5)}$ in its experimental range, up to $3\,\sigma$. The (red) arrow represents the theoretical error in the position of the lines. The (green) shaded regions are the covariance ellipses obtained combining $m_t^{MC}=173.34 \pm 0.76$ GeV and  $m_H^{exp}=125.09\pm 0.24 $ GeV; the probability of finding $m_t^{MC}$ and $m_H$ inside the inner (central, outer) ellipse is equal to $68.2\%$ ($95.5\%$, $99.7\%$). }   
\label{fig-mtmH}
\vskip .5 cm
\end{figure}

The experimental error is the one associated to the precision with which we know the input parameters at the matching scale $m_t$. The uncertainty on $m^c_t$ due to varying $\alpha_s^{(5)}$ in its $1\,\sigma$ range is $\pm 0.37$ GeV, 
while the uncertainty due to varying $m_H$ in its $1\,\sigma$ range is $\pm 0.12$ GeV.
This  can be graphically seen in fig.\,\ref{fig-mtmH}, where $m_t^c$ is displayed as a function of $m_H$ for selected values of $\alpha_s^{(5)}$; in particular, the solid line refers to its central value, while the dotted, short and long dashed lines refer to the $1\,\sigma$, $2\,\sigma$ and $3\,\sigma$ deviations respectively. In the region below (above) the line the potential is stable (metastable).

Theoretical errors comes from the approximations done in the three steps of the calculation: 
matching at $m_t$,  running from $m_t$ up to high scales, effective potential expansion. 
\begin{itemize}
\item
As already seen in section\,\ref{sec-match}, the theoretical error in the NNLO matching (scale variation and truncation) 
of $\lambda$ is equivalent to a variation in $m_H$ by about $1.6\,\sigma$ ($\pm 0.38$ GeV), which in turn means an error on $m^c_t$ of about $\pm 0.19$ GeV. The theoretical error in the NNLO matching (scale variation and truncation) of $h_t$ is equivalent to a variation in $m_t$ by about $0.3\,\sigma$, which translates into an uncertainty of about  $\pm 0.25$ GeV on $m_t^c$. Combining in quadrature, the theoretical uncertainty on $m_t^c$ due to the NNLO matching turns out to be about $\pm 0.32$ GeV: this means that the position of the straight lines in fig.\,\ref{fig-mtmH} can be shifted up or down by about $0.32$ GeV, as represented by the (red) arrow for the central value of $\alpha_s^{(5)}$.
\item
The theoretical uncertainty associated to the order of the $\beta$-functions in the RGE can be estimated by adding the order of magnitude of the unknown subsequent correction. For the NNLO, it turns out that the impact of the subsequent 4-loop correction on $m^c_t$ is of order $10^{-5}$ GeV, thus negligible.

\begin{figure}[t!]
\vskip .5cm 
 \begin{center}
\includegraphics[width=10cm]{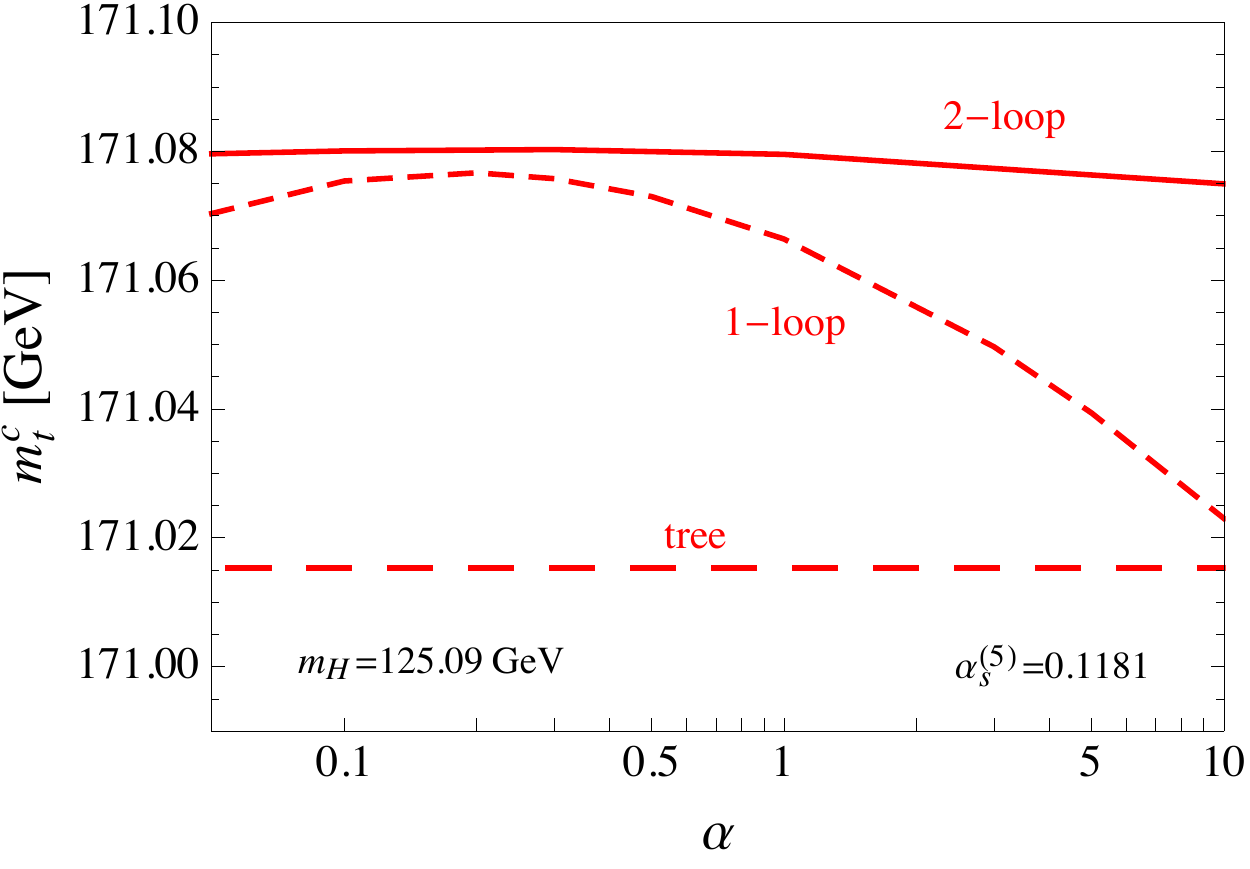} \end{center}
\caption{\baselineskip=15 pt  \small
Dependence of $m^c_t$ on $\alpha$, for increasing level of the effective potential expansion.
We choose the central values for $m_H$ and $\alpha_s^{(5)}$.  }
\label{fig-err}
\vskip .5 cm
\end{figure}
\item A way to estimate the theoretical uncertainty on $m^c_t$ associated to the order of the expansion of the effective potential, is to study its dependence on $\alpha$.  
In fig.\,\ref{fig-err} we show this dependence by varying $\alpha$ in the interval $0.1-10$, while keeping the other parameters fixed at their central values. For comparison, we display the dependence obtained by performing the calculation of the effective potential at the tree (long-dashed), 1-loop (short-dashed) and 2-loop (solid) levels.
A few comments are in order here. The renormalization scale  $\mu(t)$ depends on $\alpha$ via eq.\,(\ref{eq-alfa}) so that,
at the tree-level, the Higgs potential does not depend explicitly on $\mu(t)$, but only implicitly via the dependence of the running couplings; for this reason, in fig.\,\ref{fig-err}, the dependence on $\alpha$ for tree-level is negligible, 
and we find $m_t^{c} \approx 171.02 $ GeV.
At 1-loop the dependence on $\alpha$ is explicit but the variation is anyway small, being about $0.05$ GeV.
The 2-loop order further improves the independence on $\alpha$: we can conclude that the error on $m_t^{c}$ at the NNLO is very small, $\pm 0.005$ GeV. 

\end{itemize}

Summarizing, the theoretical error that mostly affects the NNLO calculation of $m^c_t$ is the one in the matching, while those in the truncation of the loop expansion in the $\beta$-functions and in the effective potential expansions are negligible. 

We can conclude that the result of the NNLO calculation is
\beq
m^c_t = 171.08 \pm 0.37_{\alpha_s} \pm 0.12_{m_H}  \pm 0.32_{th} \, {\rm GeV}\,,
\eeq
where the first two errors are the $1\,\sigma$ variations of $\alpha_s^{(5)}$ and $m_H$.
Our results for the value of $m^c_t$ update and improve\footnote{The calculation of the theoretical error associated to the truncation at the 2-loop order was (at our knowledge) not been shown so far.} but, modulo the doubling of the experimental error in $\alpha_s^{(5)}$, are essentially consistent with those of the most recent literature \cite{Degrassi, Bezrukov:2012sa, Alekhin:2012py, Masina:2012tz, Buttazzo:2013uya,Bednyakov:2015sca}.

\begin{figure}[t!]
\vskip 1.5cm 
 \begin{center}
\includegraphics[width=12cm]{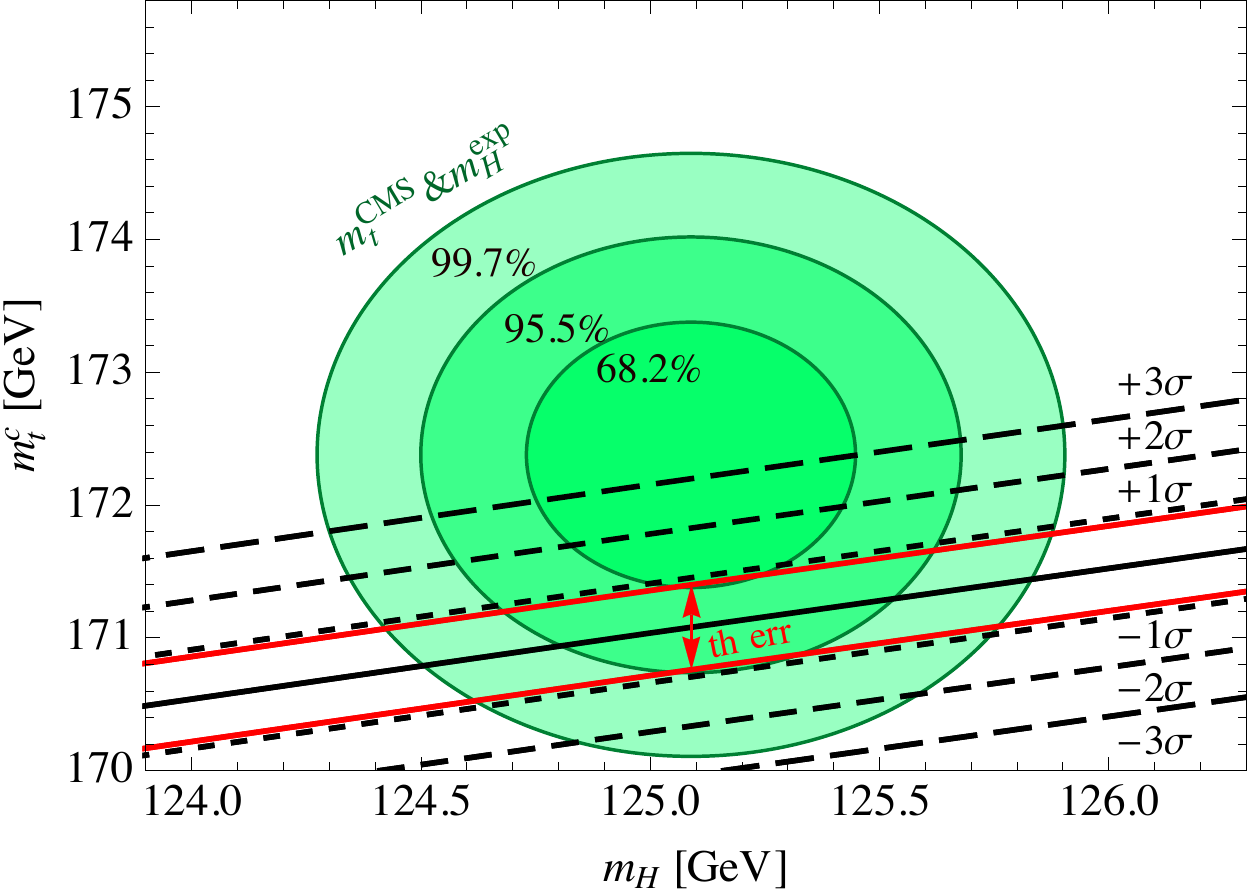}   
 \end{center}
\caption{\baselineskip=15 pt \small  
Lines for which the Higgs potential develops a second degenerate minimum at high energy. The solid line corresponds to the central value of $\alpha_s^{(5)}$; the dashed lines are obtained by varying $\alpha_s^{(5)}$ in its experimental range, up to $3\,\sigma$. The (red) arrow represents the theoretical error in the position of the lines. The (green) shaded regions are the covariance ellipses obtained combining $m_t^{\rm CMS}=172.38 \pm 0.66$ GeV \cite{CMS:2014hta} and  $m_H^{exp}=125.09\pm 0.24 $ GeV. }   
\label{fig-mtmH-CMS}
\vskip .5 cm
\end{figure}

The above value of $m_t^c$ has  to be compared with the experimental determination of the top pole mass, $m_t$.
The present value of the MC top mass is $m_t^{MC}=173.34 \pm 0.76$ GeV \cite{ATLAS:2014wva}, and would imply a difference with $m^c_t$ at the level of about  $1.7\,\sigma$. This can be graphically seen in fig.\,\ref{fig-mtmH}, where the (shaded) ellipses are the covariance ellipses for a two-valued gaussian density, obtained by combining the present experimental values of the MC top mass and Higgs mass, so that the probability of being inside the smaller, central and larger ellipse is respectively $68.2\%$, $95.5\%$ and $99.7\%$. However, as discussed in section\,\ref{sec-match}, the uncertainty in the identification between the pole and MC top mass is currently estimated to be of order $200$ MeV \cite{Nason:2016tiy, Corcella:2015kth} (or even $1$ GeV for the most conservative groups \cite{Moch:2014tta}): this would further reduce the difference, say at the $1.5\,\sigma$ level. 
Statistically speaking, a tension at the $1.5\,\sigma$ level supports stability and metastability at the $14\%$ and $86\%$ C.L. respectively. Physically speaking, in our opinion, it is even too strong to use the term `tension' when speaking of a $1.5\,\sigma$ deviation.   

As a result, the updated calculation of the experimental and theoretical uncertainties on $m_t^c$, in addition to the uncertainty in the identification of the MC and pole top masses, lead us to conclude that the configuration with two degenerate vacua is at present compatible with the experimental data.

Our conclusions agree with those of ref. \cite{Bednyakov:2015sca}. Previous claims that stability is disfavored at more than the $95\%$ level \cite{Degrassi, Buttazzo:2013uya} are due, in our opinion, to the previous underestimation of two uncertainties -- the experimental one in the determination of $\alpha_s^{(5)}$ and the theoretical one in the identification of the MC and pole top masses --, together with a less conservative interpretation (with respect to ours) of the significance of the results.

It is clear that, in order to discriminate in a robust way between stability and metastability, it would be crucial to reduce the experimental uncertainties in both $m_t$ and $\alpha_s^{(5)}$. A reduction of the theoretical error in the matching would also be welcome.

As a final remark, notice that a recent measurement of the top pole mass by the CMS Collaboration is $m_t^{\rm CMS}= 172.38 \pm 0.66$ GeV \cite{CMS:2014hta}. The shaded ellipses in this case would change as shown in fig.\,\ref{fig-mtmH-CMS}, and the discrepancy with $m^c_t$ would thus further decrease, at less than $1\,\sigma$. It will be very interesting to see if such low value will be confirmed by future LHC data.

\section{Inflection point}
\label{sec-inf}

We now turn to the inflection point configuration assuming, as usual, that the potential at the electroweak minimum is zero. 
Such configuration could be relevant for the class of models of primordial inflation based on a shallow false minimum \cite{MasinaHiggsmass,Masinatop,Masinahybrid,Masinaupgrade}.
In particular the highness of the effective potential at an inflection point, let us call it $\bar V_i$, could be directly linked to the ratio of the scalar to tensor modes of primordial perturbations, $r$, via the relation
\beq
\bar V_{i}  = \frac{3 \pi^2}{2} \, r \, A_s \, ,
\label{eq-r}
\eeq
where the amplitude of scalar perturbations is $A_s =2.2 \times 10^{-9} $ \cite{Ade:2015xua}.
It is thus relevant for models of primordial inflation to assess the size of the experimental and theoretical errors in the calculation of $\bar V_i$.

\begin{figure}[t!]
\vskip .5cm 
 \begin{center}
  \includegraphics[width=12cm]{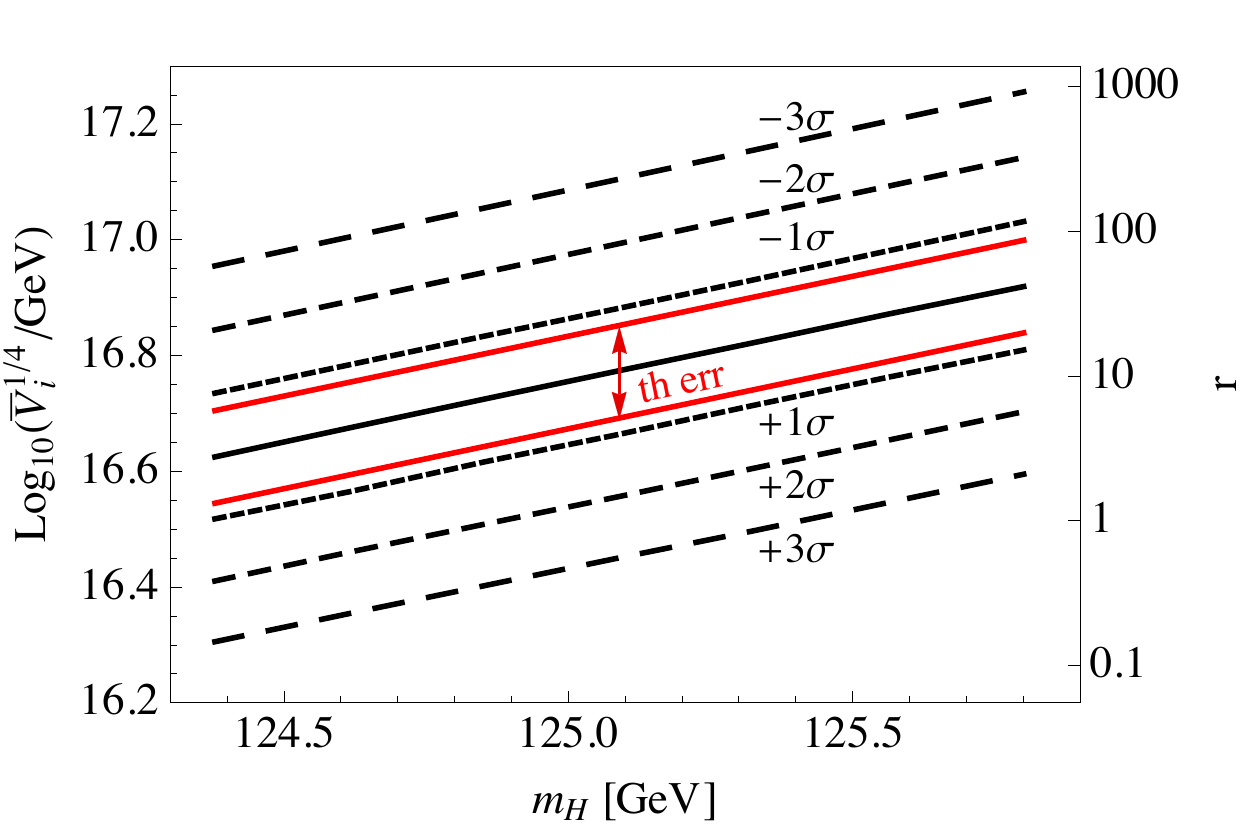}   
 \end{center}
\caption{\baselineskip=15 pt \small 
Dependence of $\bar V_i^{1/4}$ on $m_H$ for fixed values of $\alpha_s^{(5)}$. The (red) arrow and solid lines show the theoretical error due to the matching of $\lambda$ (positive contribution to $\lambda$ for the upper line, negative for the lower one). The right vertical axis displays the associated value of the tensor-to-scalar ratio $r$, according to eq.\,(\ref{eq-r}).}
\label{fig-Veff}
\vskip .1 cm
\end{figure}

Experimental uncertainties on $\bar V_i$ can be estimated as follows. We let $m_H$ vary in its $3\,\sigma$ experimental range and, for fixed values of $\alpha_s^{(5)}$,
we determine $m^i_t$, the value of the top mass needed to have an inflection point (which is so close to $m^c_t$ that one can 
read it from the stability line of fig.\,\ref{fig-mtmH}). We then evaluate the effective potential at this point, $\bar V_i$. 
The result is displayed in fig.\,\ref{fig-Veff}: one can see that $\bar V_i^{1/4}$ spans one order of magnitude, as it varies from $2\times 10^{16}$ GeV up to $2 \times 10^{17}$ GeV, for decreasing values of $\alpha_s^{(5)}$; the dependence on $m_H$ is less dramatic but still relevant, being about $50\%$.

\begin{figure}[t!]
\vskip .5cm 
 \begin{center}
  \includegraphics[width=11cm]{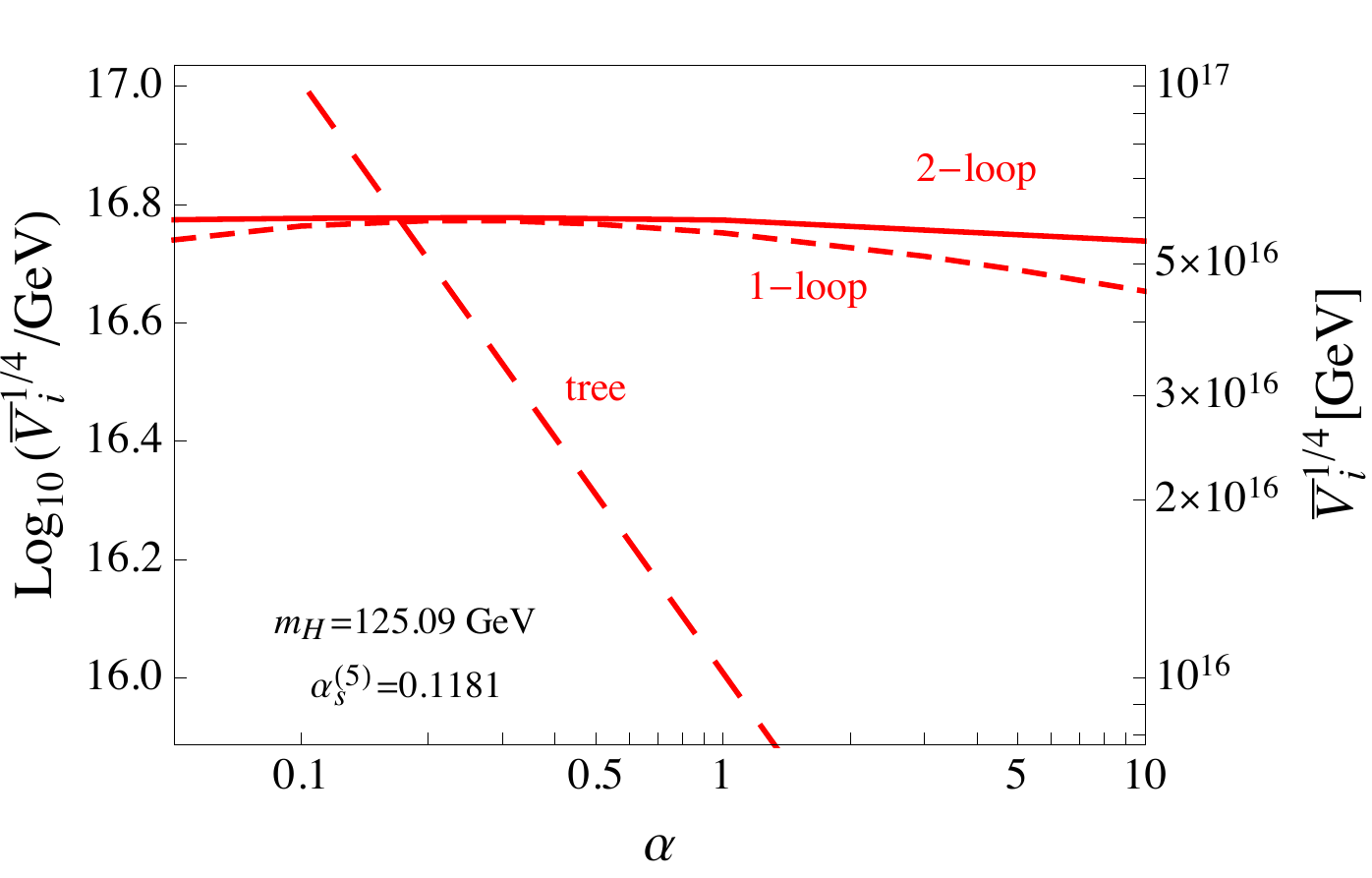}   
 \end{center}
\caption{\baselineskip=15 pt \small
Dependence of $\bar V_{i}^{1/4}$ on $\alpha$. For definiteness, $\alpha_s^{(5)}$ and $m_H$ are assigned to their central values.}
\label{fig-bVi}
\vskip .1 cm
\end{figure}

We divide the theoretical errors in three categories, as done before.
\begin{itemize}
\item 
Theoretical errors associated to matching of $\lambda$ at NNLO, have an impact on $m_t^c$ of about $\pm 0.19$ GeV (see fig.\,\ref{fig-mtmH} in the previous section) and an impact on $\log_{10}(\bar V_i^{1/4}/{\rm GeV})$ of about $\pm 0.08$, namely a $20\%$ variation of $\bar V_i^{1/4}$. As for the NNLO matching of $h_t$, the impact on $m_t^c$ is of about $\pm 0.25$ GeV, but there is no significant impact on $\bar V_i$. As a consequence of the theoretical error in the matching, the lines in fig.\,\ref{fig-Veff} could be shifted up or down by about $0.08$; for the central value of $\alpha_s^{(5)}$, this is represented in fig.\,\ref{fig-Veff} by the (red) arrow and solid lines. The theoretical error due to the NNLO matching is thus slightly smaller than the experimental error due to the $1\,\sigma$ variation of $\alpha^{(5)}$.

\item
The order of magnitude of the theoretical errors associated to the $\beta$-functions at NNLO can be estimated by studying the impact of the subsequent correction; it turns out that $\bar V_{i}^{1/4}$ changes at the per mille level. Such error is thus negligible.

\item
The theoretical uncertainty associated to the fact that we truncate the effective potential at 2-loop can be estimated by studying the dependence of $\bar V_{i}$ on $\alpha$. We fix $\alpha_s^{(5)}$ and $m_H$ at their central values and display in fig.\,\ref{fig-bVi} the resulting value of $\bar V_{i}^{1/4}$ by means of the solid line;
 for comparison, we display also the dependences obtained at tree\footnote{We checked that our results for the tree level with $\alpha=1$ agree with those obtained in refs.\,\cite{Masina:2012tz, Masinaupgrade}. In these works, the potential was indeed calculated only at the tree-level, without providing any estimate for the theoretical uncertainty because of neglecting higher loops.} (long-dashed) and 1-loop (short-dashed) levels.
Notice that the dependence of $\bar V_{i}$ on $\alpha$ at the tree-level is implicit\footnote{Indeed, $V_{\text{eff}} \propto \lambda\left( \ln( \alpha  \phi / m_t ) \right)$.}, but significant: the tree-level calculation of $\bar V_{i}^{1/4}$ is uncertain by more than one order of magnitude. The 1-loop corrections flattens the dependence on $\alpha$ so that the uncertainty on $\bar V_i^{1/4}$ gets reduced down to about $20\%$; this uncertainty is comparable to the theoretical one due to the matching. 
The 2-loop correction further flattens the dependence on $\alpha$ and allows to estimate $\bar V_{i}^{1/4}$ with a $5\%$ precision.  
\end{itemize}

Summarizing, the theoretical error that mostly affects the calculation of $\bar V_{i}$ at the NNLO is the one associated to the matching.
The theoretical error in the truncation of the effective potential is smaller than the theoretical error in the matching only including the 2-loop correction to the effective potential.

We can conclude that the result of the NNLO calculation is
\beq
\log_{10} {\bar V}_{i}^{1/4} =  16.77  \pm 0.11_{\, \alpha_s} \pm 0.05_{m_H}  \pm 0.08_{th}  \,,
\label{eq-V14}
\eeq
where the first two errors refer the $1\,\sigma$ variations of $\alpha_s^{(5)}$ and $m_H$ respectively, 
while the theoretical error is essentially dominated by the one in the matching.

As anticipated, a precise determination of $\bar V_i$ is important for models of inflation based on the idea of a shallow false minimum \cite{MasinaHiggsmass,Masinatop,Masinahybrid,Masinaupgrade} as, in these models, $\bar V_i$ and $r$ are linked via eq.\,(\ref{eq-r}). In view of such application, the right axis of fig.\,\ref{fig-Veff} reports the corresponding value of $r$. Notice that the dependence of $r$ on $\alpha_s^{(5)}$ and $m_H$ is extremely strong: when the latter are varied in their $3\,\sigma$ range, $r$ spans about four orders of magnitude, from $0.1$ to $1000$. In addition, the theoretical error in the matching implies an uncertainty on $r$ by a factor of about $2$.

According to the 2015 analysis of the Planck Collaboration, the present upper bound on the tensor to scalar ratio is $r <0.12$ at $95\%$ C.L. \cite{Ade:2015lrj}, as also confirmed by the recent joint analysis with the BICEP2 Collaboration \cite{Ade:2015tva}. 
Due to eq.\,(\ref{eq-r}), this would translate into to the bound $\log_{10} {\bar V}_i^{1/4} < 16.28$ 
at $95\%$ C.L.; this implies a tension with eq.\,(\ref{eq-V14}) at about $3\,\sigma$.

\begin{figure}[t!]
\vskip .5cm 
 \begin{center}
  \includegraphics[width=11cm]{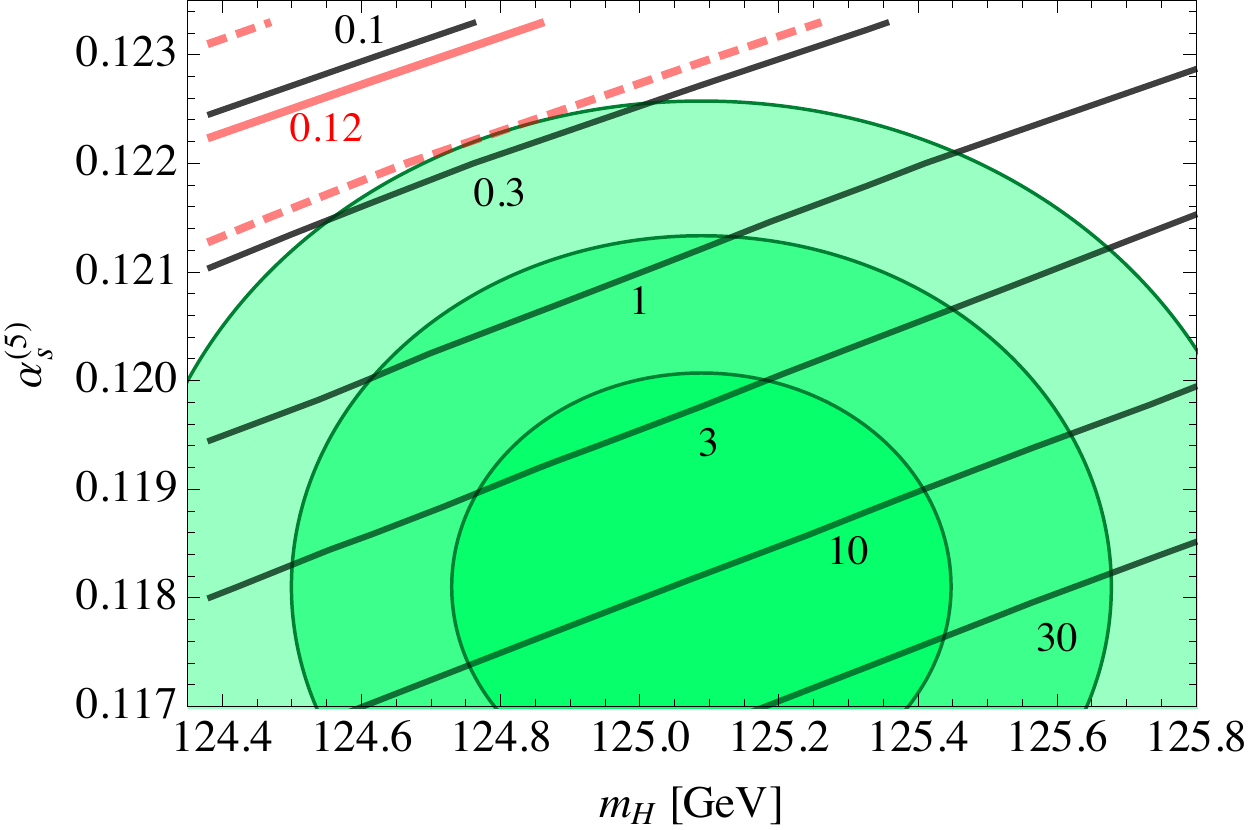}   
 \end{center}
\caption{\baselineskip=15 pt \small Contour levels of $r$ in the plane $(m_H, \alpha_s^{(5)})$. The theoretical uncertainty corresponding to $r=0.12$ is shown by means of the (red) dashed lines. The shaded regions are the covariance ellipses indicating that the probability of finding the experimental values of $m_H$ and $ \alpha_s^{(5)}$  inside the ellipses are respectively $68.2\%, 95.4\%,99.7\% $.
 }
\label{fig-r}
\vskip .1 cm
\end{figure}

This can be graphically seen in fig.\,\ref{fig-r}, where the contour levels of $r$ in the plane $(m_H, \alpha_s^{(5)})$ are shown.
Even invoking the uncertainty due to the matching (red-dashed lines), a value for $r$ as small as $0.12$ (red-solid line), could be obtained only with $\alpha_s^{(5)}$ in its upper $3\,\sigma$ range, while the values of $m_H$, and also $m_t$  (see fig.\,\ref{fig-mtmH}), could stay inside their $1\,\sigma$ interval.

A value for $r$ close to $0.2$, as claimed by BICEP2 in 2013 \cite{Bicep}, 
would have been compatible at about $2\,\sigma$ with a model based on a shallow false minimum.
With the inclusion of the theoretical error and the slight changes in the central values of the input parameters,
the present calculation superseeds the results obtained in the previous paper \cite{Masinaupgrade}, where the calculation of the effective potential was done only at tree-level (due to the yet unsolved problem of the inclusion of the 1-loop correction when $\lambda$ turns negative \cite{Andreassen:2014gha}).

The present results have to be compared with those of the most recent detailed analysis on the inflection point configuration, performed by Ballesteros et al. \cite{Ballesteros:2015iua}. The latter work includes the 2-loop correction to the Higgs effective potential but, at our understanding, do not include the theoretical error in the matching (which is actually the dominant one according to us). This may justify their stronger exclusion of the inflection point configuration. Apart from this detail, the results of the two analysis are in substantial agreement. The even stronger exclusion of the inflection point configuration found in ref. \cite{Notari:2014noa}, might be also due to an underestimation of the theoretical errors, together with the previous underestimation of the experimental uncertainty on $\alpha_s^{(5)}$.

\section{Gravitational contribution: effects of physics at $M_{P}$}
\label{sec-grav}

Through this work we assumed the SM to be valid all the way up to the Planck scale. Near the cutoff of the theory, large Planckian effects are possible, but without a satisfactory comprehension of quantum gravity effects (with a reliable UV completion of the theory) there is no hope to calculate them. The usual approach in this sense is to consider a tower of a non-renormalizable operators suppressed by the cutoff in an effective theory scenario below $M_{P}$\,\cite{Branchina:2013jra , Branchina:2014usa, Branchina:2015nda, Loebbert:2015eea, Lalak:2015bwa, Espinosa:2015qea}, leading to a modification of the SM Higgs potential:
\begin{equation} \label{higher}
V(\phi)=\frac{\lambda}{24}\phi^{4}+\frac{\lambda_{6}}{6}\frac{\phi^{6}}{M_{P}^{2}}+\frac{\lambda_{8}}{8}\frac{\phi^{8}}{M_{P}^{4}}+\mathcal{O}\left(\frac{\phi^{10}}{M_{P}^{6}}\right)\,.
\end{equation}
Without any protecting symmetry, these corrections have to be taken into account. In this way, assuming order one values for the new unknown couplings $\lambda_{6}$ and $\lambda_{8}$, treated as free parameters, it is in principle possible to estimate the impact of gravitational physics. The effects of these higher-order operators turn out to be heavily dependent on the choice of the free couplings towards both stability and instability: it is not clear why gravitational physics should make the potential more unstable or vice versa.
The approach of eq.\,(\ref{higher}) has however raised some concerns \cite{Espinosa:2015kwx}, as the method is based on an effective theory expansion that breaks down when $\phi\sim M_{P}$. The use of an effective theory close to its cutoff might not be fully reliable.

\section{Conclusions}
\label{sec-concl}

We studied in detail gauge-independent observables associated with two interesting stationary configurations of the SM Higgs potential extrapolated at the NNLO:  
i) the value of the top mass ensuring stability of the SM electroweak minimum, and ii) the value of the Higgs potential at a rising inflection point. 
We reappraised in a critical way the experimental and theoretical uncertainties plaguing their determination. 

Considering the updated value of the experimental error on $\alpha_s^{(5)}$, the issue of the determination of the top pole mass from the MC one, and the theoretical uncertainty associated to the matching, we find that stability of the SM is compatible with the present data at the level of $1.5\,\sigma$. 
Stability of the SM Higgs potential is thus, in our opinion, a viable possibility. 
In order to robustly discriminate between stability and metastability, higher precision measurements of the top quark pole mass 
and of $\alpha_s^{(5)}$ would be needed. For the time being, we can wonder whether the fact that the Higgs potential is so close to a configuration with two degenerate vacua is telling us something deep \cite{Espinosa:2015qea}. 

As for the configuration of a rising inflection point, we find that, despite the large theoretical error plaguing the value of the Higgs potential at the inflection point, application of such configuration to models of primordial inflation based on a shallow false minimum displays a $3\,\sigma$ tension with the recent bounds on the tensor-to-scalar ratio of cosmological perturbations. 
The tension is essentially due to the value of $\alpha_s^{(5)}$, instead of the top or Higgs masses. Hence, if $\alpha_s^{(5)}$ will turn out to be in its present $3\,\sigma$ range, such models will be rescued; otherwise, modifications due to new physics will be necessarily introduced \cite{EliasMiro:2012ay, Ballesteros:2015iua}.

\section*{\large Acknowledgements}

This work was supported by the Italian Ministero dell'Istruzione, Universit\`a e Ricerca (MIUR) and Istituto Nazionale di Fisica Nucleare (INFN) through the "Theoretical Astroparticle Physics" research projects. The PhD grant of G.I. is supported by the Agenzia Spaziale Italiana (ASI). We warmly thanks Gennaro Corcella, Jos\'e Ram\'on Espinosa, Guido Martinelli and Alberto Salvio for very useful discussions. I.M. thanks the CERN Theory Department for kind hospitality and support during the completion of this work.

\bibliographystyle{elsarticle-num} 
\bibliography{bib} 

\begin{thebibliography}{10}
\expandafter\ifx\csname url\endcsname\relax
  \def\url#1{\texttt{#1}}\fi
\expandafter\ifx\csname urlprefix\endcsname\relax\def\urlprefix{URL }\fi
\expandafter\ifx\csname href\endcsname\relax
  \def\href#1#2{#2} \def\path#1{#1}\fi

\bibitem{Hung:1979dn}
P.~Q. Hung, {Vacuum Instability and New Constraints on Fermion Masses}, Phys.
  Rev. Lett. 42 (1979) 873.
\newblock \href {http://dx.doi.org/10.1103/PhysRevLett.42.873}
  {\path{doi:10.1103/PhysRevLett.42.873}}.

\bibitem{Cabibbo:1979ay}
N.~Cabibbo, L.~Maiani, G.~Parisi, R.~Petronzio, {Bounds on the Fermions and
  Higgs Boson Masses in Grand Unified Theories}, Nucl. Phys. B158 (1979)
  295--305.
\newblock \href {http://dx.doi.org/10.1016/0550-3213(79)90167-6}
  {\path{doi:10.1016/0550-3213(79)90167-6}}.

\bibitem{Froggatt:1995rt}
C.~D. Froggatt, H.~B. Nielsen, {Standard model criticality prediction: Top mass
  173 $\pm$ 5 GeV and Higgs mass 135 $\pm$ 9 GeV}, Phys. Lett. B368 (1996)
  96--102.
\newblock \href {http://arxiv.org/abs/hep-ph/9511371}
  {\path{arXiv:hep-ph/9511371}}, \href
  {http://dx.doi.org/10.1016/0370-2693(95)01480-2}
  {\path{doi:10.1016/0370-2693(95)01480-2}}.

\bibitem{MasinaHiggsmass}
I.~Masina, A.~Notari, {The Higgs mass range from Standard Model false vacuum
  Inflation in scalar-tensor gravity}, Phys.Rev. D85 (2012) 123506.
\newblock \href {http://arxiv.org/abs/1112.2659} {\path{arXiv:1112.2659}},
  \href {http://dx.doi.org/10.1103/PhysRevD.85.123506}
  {\path{doi:10.1103/PhysRevD.85.123506}}.

\bibitem{atlas}
G.~Aad, et~al., {Observation of a new particle in the search for the Standard
  Model Higgs boson with the ATLAS detector at the LHC}, Physics Letters B 716
  (2012) 1 -- 29.
\newblock \href
  {http://dx.doi.org/http://dx.doi.org/10.1016/j.physletb.2012.08.020}
  {\path{doi:http://dx.doi.org/10.1016/j.physletb.2012.08.020}}.

\bibitem{cms}
S.~Chatrchyan, et~al., {Observation of a new boson at a mass of 125 GeV with
  the CMS experiment at the LHC}, Physics Letters B 716 (2012) 30--61.
\newblock \href
  {http://dx.doi.org/http://dx.doi.org/10.1016/j.physletb.2012.08.021}
  {\path{doi:http://dx.doi.org/10.1016/j.physletb.2012.08.021}}.

\bibitem{Degrassi}
G.~Degrassi, S.~Di~Vita, J.~Elias-Miro, J.~R. Espinosa, G.~F. Giudice, et~al.,
  {Higgs mass and vacuum stability in the Standard Model at NNLO}, JHEP 1208
  (2012) 098.
\newblock \href {http://arxiv.org/abs/1205.6497} {\path{arXiv:1205.6497}},
  \href {http://dx.doi.org/10.1007/JHEP08(2012)098}
  {\path{doi:10.1007/JHEP08(2012)098}}.

\bibitem{Bezrukov:2012sa}
F.~Bezrukov, M.~{\relax Yu}. Kalmykov, B.~A. Kniehl, M.~Shaposhnikov, {Higgs
  Boson Mass and New Physics}, JHEP 10 (2012) 140.
\newblock \href {http://arxiv.org/abs/1205.2893} {\path{arXiv:1205.2893}},
  \href {http://dx.doi.org/10.1007/JHEP10(2012)140}
  {\path{doi:10.1007/JHEP10(2012)140}}.

\bibitem{Buttazzo:2013uya}
D.~Buttazzo, G.~Degrassi, P.~P. Giardino, G.~F. Giudice, F.~Sala, A.~Salvio,
  A.~Strumia, {Investigating the near-criticality of the Higgs boson}, JHEP 12
  (2013) 089.
\newblock \href {http://arxiv.org/abs/1307.3536} {\path{arXiv:1307.3536}},
  \href {http://dx.doi.org/10.1007/JHEP12(2013)089}
  {\path{doi:10.1007/JHEP12(2013)089}}.

\bibitem{Bednyakov:2015sca}
A.~V. Bednyakov, B.~A. Kniehl, A.~F. Pikelner, O.~L. Veretin, {Stability of the
  Electroweak Vacuum: Gauge Independence and Advanced Precision}, Phys. Rev.
  Lett. 115~(20) (2015) 201802.
\newblock \href {http://arxiv.org/abs/1507.08833} {\path{arXiv:1507.08833}},
  \href {http://dx.doi.org/10.1103/PhysRevLett.115.201802}
  {\path{doi:10.1103/PhysRevLett.115.201802}}.

\bibitem{DiLuzio:2014bua}
L.~Di~Luzio, L.~Mihaila, {On the gauge dependence of the Standard Model vacuum
  instability scale}, JHEP 06 (2014) 079.
\newblock \href {http://arxiv.org/abs/1404.7450} {\path{arXiv:1404.7450}},
  \href {http://dx.doi.org/10.1007/JHEP06(2014)079}
  {\path{doi:10.1007/JHEP06(2014)079}}.

\bibitem{Espinosa:2015qea}
J.~R. Espinosa, G.~F. Giudice, E.~Morgante, A.~Riotto, L.~Senatore, A.~Strumia,
  N.~Tetradis, {The cosmological Higgstory of the vacuum instability}, JHEP 09
  (2015) 174.
\newblock \href {http://arxiv.org/abs/1505.04825} {\path{arXiv:1505.04825}},
  \href {http://dx.doi.org/10.1007/JHEP09(2015)174}
  {\path{doi:10.1007/JHEP09(2015)174}}.

\bibitem{Andreassen:2014eha}
A.~Andreassen, W.~Frost, M.~D. Schwartz, {Consistent Use of Effective
  Potentials}, Phys. Rev. D91~(1) (2015) 016009.
\newblock \href {http://arxiv.org/abs/1408.0287} {\path{arXiv:1408.0287}},
  \href {http://dx.doi.org/10.1103/PhysRevD.91.016009}
  {\path{doi:10.1103/PhysRevD.91.016009}}.

\bibitem{Andreassen:2014gha}
A.~Andreassen, W.~Frost, M.~D. Schwartz, {Consistent Use of the Standard Model
  Effective Potential}, Phys. Rev. Lett. 113~(24) (2014) 241801.
\newblock \href {http://arxiv.org/abs/1408.0292} {\path{arXiv:1408.0292}},
  \href {http://dx.doi.org/10.1103/PhysRevLett.113.241801}
  {\path{doi:10.1103/PhysRevLett.113.241801}}.

\bibitem{Aad:2015zhl}
G.~Aad, et~al., {Combined Measurement of the Higgs Boson Mass in $pp$
  Collisions at $\sqrt{s}=7$ and 8 TeV with the ATLAS and CMS Experiments},
  Phys. Rev. Lett. 114 (2015) 191803.
\newblock \href {http://arxiv.org/abs/1503.07589} {\path{arXiv:1503.07589}},
  \href {http://dx.doi.org/10.1103/PhysRevLett.114.191803}
  {\path{doi:10.1103/PhysRevLett.114.191803}}.

\bibitem{Agashe:2015}
S.~Bethke, G.~Dissertori, G.~P. Salam,
  \href{{http://pdg.lbl.gov/2015/reviews/rpp2015-rev-qcd.pdf}}{{Particle Data
  Group review on Quantum Chromodynamics, revised version of September 2015.
  }}.
\newline\urlprefix\url{{http://pdg.lbl.gov/2015/reviews/rpp2015-rev-qcd.pdf}}

\bibitem{Moch:2014lka}
S.~Moch, {Precision determination of the top-quark mass}, PoS LL2014 (2014)
  054.
\newblock \href {http://arxiv.org/abs/1408.6080} {\path{arXiv:1408.6080}}.

\bibitem{Nason:2016tiy}
P.~Nason, {Theory Summary. }\href {http://arxiv.org/abs/1602.00443}
  {\path{arXiv:1602.00443}}.

\bibitem{Corcella:2015kth}
G.~Corcella, {Interpretation of the top-quark mass measurements: a theory
  overview}, in: {8th International Workshop on Top Quark Physics (TOP2015)
  Ischia, NA, Italy, September 14-18, 2015}, 2015.
\newblock \href {http://arxiv.org/abs/1511.08429} {\path{arXiv:1511.08429}}.

\bibitem{ATLAS:2014wva}
G.~Aad, et~al., {First combination of Tevatron and LHC measurements of the
  top-quark mass. }\href {http://arxiv.org/abs/1403.4427}
  {\path{arXiv:1403.4427}}.

\bibitem{Alekhin:2012py}
S.~Alekhin, A.~Djouadi, S.~Moch, {The top quark and Higgs boson masses and the
  stability of the electroweak vacuum}, Phys. Lett. B716 (2012) 214--219.
\newblock \href {http://arxiv.org/abs/1207.0980} {\path{arXiv:1207.0980}},
  \href {http://dx.doi.org/10.1016/j.physletb.2012.08.024}
  {\path{doi:10.1016/j.physletb.2012.08.024}}.

\bibitem{Masina:2012tz}
I.~Masina, {Higgs boson and top quark masses as tests of electroweak vacuum
  stability}, Phys. Rev. D87~(5) (2013) 053001.
\newblock \href {http://arxiv.org/abs/1209.0393} {\path{arXiv:1209.0393}},
  \href {http://dx.doi.org/10.1103/PhysRevD.87.053001}
  {\path{doi:10.1103/PhysRevD.87.053001}}.

\bibitem{Bezrukov:2014ina}
F.~Bezrukov, M.~Shaposhnikov, {Why should we care about the top quark Yukawa
  coupling?}, J. Exp. Theor. Phys. 120 (2015) 335--343, [Zh. Eksp. Teor.
  Fiz.147,389(2015)].
\newblock \href {http://arxiv.org/abs/1411.1923} {\path{arXiv:1411.1923}},
  \href {http://dx.doi.org/10.1134/S1063776115030152}
  {\path{doi:10.1134/S1063776115030152}}.

\bibitem{Notari:2014noa}
A.~Notari, {Higgs Mass and Gravity Waves in Standard Model False Vacuum
  Inflation}, Phys. Rev. D91 (2015) 063527.
\newblock \href {http://arxiv.org/abs/1405.6943} {\path{arXiv:1405.6943}},
  \href {http://dx.doi.org/10.1103/PhysRevD.91.063527}
  {\path{doi:10.1103/PhysRevD.91.063527}}.

\bibitem{Ballesteros:2015iua}
G.~Ballesteros, C.~Tamarit, {Higgs portal valleys, stability and inflation},
  JHEP 09 (2015) 210.
\newblock \href {http://arxiv.org/abs/1505.07476} {\path{arXiv:1505.07476}},
  \href {http://dx.doi.org/10.1007/JHEP09(2015)210}
  {\path{doi:10.1007/JHEP09(2015)210}}.

\bibitem{Agashe:2014kda}
K.~A. Olive, et~al., {Review of Particle Physics}, Chin. Phys. C38 (2014)
  090001.
\newblock \href {http://dx.doi.org/10.1088/1674-1137/38/9/090001}
  {\path{doi:10.1088/1674-1137/38/9/090001}}.

\bibitem{Coleman}
S.~Coleman, E.~Weinberg, {Radiative Corrections as the Origin of Spontaneous
  Symmetry Breaking}, Phys. Rev. D 7 (1973) 1888--1910.
\newblock \href {http://dx.doi.org/10.1103/PhysRevD.7.1888}
  {\path{doi:10.1103/PhysRevD.7.1888}}.

\bibitem{PDG2012}
J.~Beringer, et~al., Review of particle physics*, Phys. Rev. D 86 (2012)
  010001.
\newblock \href {http://dx.doi.org/10.1103/PhysRevD.86.010001}
  {\path{doi:10.1103/PhysRevD.86.010001}}.

\bibitem{Moch:2014tta}
S.~Moch, et~al., {High precision fundamental constants at the TeV scale. }\href
  {http://arxiv.org/abs/1405.4781} {\path{arXiv:1405.4781}}.

\bibitem{Hoang:2008xm}
A.~H. Hoang, I.~W. Stewart, {Top Mass Measurements from Jets and the Tevatron
  Top-Quark Mass}, Nucl. Phys. Proc. Suppl. 185 (2008) 220--226.
\newblock \href {http://arxiv.org/abs/0808.0222} {\path{arXiv:0808.0222}},
  \href {http://dx.doi.org/10.1016/j.nuclphysbps.2008.10.028}
  {\path{doi:10.1016/j.nuclphysbps.2008.10.028}}.

\bibitem{Fleming:2007xt}
S.~Fleming, A.~H. Hoang, S.~Mantry, I.~W. Stewart, {Top Jets in the Peak
  Region: Factorization Analysis with NLL Resummation}, Phys. Rev. D77 (2008)
  114003.
\newblock \href {http://arxiv.org/abs/0711.2079} {\path{arXiv:0711.2079}},
  \href {http://dx.doi.org/10.1103/PhysRevD.77.114003}
  {\path{doi:10.1103/PhysRevD.77.114003}}.

\bibitem{Marquard:2015qpa}
P.~Marquard, A.~V. Smirnov, V.~A. Smirnov, M.~Steinhauser, {Quark Mass
  Relations to Four-Loop Order in Perturbative QCD}, Phys. Rev. Lett. 114~(14)
  (2015) 142002.
\newblock \href {http://arxiv.org/abs/1502.01030} {\path{arXiv:1502.01030}},
  \href {http://dx.doi.org/10.1103/PhysRevLett.114.142002}
  {\path{doi:10.1103/PhysRevLett.114.142002}}.

\bibitem{Kataev:2015gvt}
A.~L. Kataev, V.~S. Molokoedov, {On the flavour dependence of the
  $\mathcal{O}(\alpha_s^4)$ correction to the relation between running and pole
  heavy quark masses. }\href {http://arxiv.org/abs/1511.06898}
  {\path{arXiv:1511.06898}}.

\bibitem{Ford:1992}
C.~Ford, I.~Jack, D.~R.~T. Jones, {The Standard model effective potential at
  two loops}, Nucl. Phys. B387 (1992) 373--390, [Erratum: Nucl.
  Phys.B504,551(1997)].
\newblock \href {http://arxiv.org/abs/hep-ph/0111190}
  {\path{arXiv:hep-ph/0111190}}, \href
  {http://dx.doi.org/10.1016/0550-3213(92)90165-8}
  {\path{doi:10.1016/0550-3213(92)90165-8}}.

\bibitem{Mihaila:2012}
L.~N. Mihaila, J.~Salomon, M.~Steinhauser, {Gauge Coupling $\beta$-functions in
  the Standard Model to Three Loops}, Phys. Rev. Lett. 108 (2012) 151602.
\newblock \href {http://arxiv.org/abs/1201.5868} {\path{arXiv:1201.5868}},
  \href {http://dx.doi.org/10.1103/PhysRevLett.108.151602}
  {\path{doi:10.1103/PhysRevLett.108.151602}}.

\bibitem{Mihaila1}
L.~N. Mihaila, J.~Salomon, M.~Steinhauser, {Renormalization constants and
  $\beta$-functions for the gauge couplings of the Standard Model to three-loop
  order}, Phys. Rev. D86 (2012) 096008.
\newblock \href {http://arxiv.org/abs/1208.3357} {\path{arXiv:1208.3357}},
  \href {http://dx.doi.org/10.1103/PhysRevD.86.096008}
  {\path{doi:10.1103/PhysRevD.86.096008}}.

\bibitem{ChetyrkinZoller}
K.~G. Chetyrkin, M.~F. Zoller, {Three-loop $\beta$-functions for top-Yukawa and
  the Higgs self-interaction in the Standard Model}, JHEP 06 (2012) 033.
\newblock \href {http://arxiv.org/abs/1205.2892} {\path{arXiv:1205.2892}},
  \href {http://dx.doi.org/10.1007/JHEP06(2012)033}
  {\path{doi:10.1007/JHEP06(2012)033}}.

\bibitem{Chetyrkin:2013}
K.~G. Chetyrkin, M.~F. Zoller, {$\beta$-function for the Higgs self-interaction
  in the Standard Model at three-loop level}, JHEP 04 (2013) 091, [Erratum:
  JHEP09,155(2013)].
\newblock \href {http://arxiv.org/abs/1303.2890} {\path{arXiv:1303.2890}},
  \href {http://dx.doi.org/10.1007/JHEP04(2013)091, 10.1007/JHEP09(2013)155}
  {\path{doi:10.1007/JHEP04(2013)091, 10.1007/JHEP09(2013)155}}.

\bibitem{BednyakovPikelnerVelizhanin}
A.~V. Bednyakov, A.~F. Pikelner, V.~N. Velizhanin, {Higgs self-coupling
  $\beta$-function in the Standard Model at three loops}, Nucl. Phys. B875
  (2013) 552--565.
\newblock \href {http://arxiv.org/abs/1303.4364} {\path{arXiv:1303.4364}},
  \href {http://dx.doi.org/10.1016/j.nuclphysb.2013.07.015}
  {\path{doi:10.1016/j.nuclphysb.2013.07.015}}.

\bibitem{BednyakovPikelnerVelizhanin1}
A.~V. Bednyakov, A.~F. Pikelner, V.~N. Velizhanin, {Yukawa coupling
  $\beta$-functions in the Standard Model at three loops}, Phys. Lett. B722
  (2013) 336--340.
\newblock \href {http://arxiv.org/abs/1212.6829} {\path{arXiv:1212.6829}},
  \href {http://dx.doi.org/10.1016/j.physletb.2013.04.038}
  {\path{doi:10.1016/j.physletb.2013.04.038}}.

\bibitem{Bednyakov:2013}
A.~V. Bednyakov, A.~F. Pikelner, V.~N. Velizhanin, {Three-loop Higgs
  self-coupling $\beta$-function in the Standard Model with complex Yukawa
  matrices}, Nucl. Phys. B879 (2014) 256--267.
\newblock \href {http://arxiv.org/abs/1310.3806} {\path{arXiv:1310.3806}},
  \href {http://dx.doi.org/10.1016/j.nuclphysb.2013.12.012}
  {\path{doi:10.1016/j.nuclphysb.2013.12.012}}.

\bibitem{Bednyakov:2014}
A.~V. Bednyakov, A.~F. Pikelner, V.~N. Velizhanin, {Three-loop SM
  $\beta$-functions for matrix Yukawa couplings}, Phys. Lett. B737 (2014)
  129--134.
\newblock \href {http://arxiv.org/abs/1406.7171} {\path{arXiv:1406.7171}},
  \href {http://dx.doi.org/10.1016/j.physletb.2014.08.049}
  {\path{doi:10.1016/j.physletb.2014.08.049}}.

\bibitem{Zoller:2015tha}
M.~F. Zoller, {Top-Yukawa effects on the $\beta$-function of the strong
  coupling in the SM at four-loop level}, JHEP 02 (2016) 095.
\newblock \href {http://arxiv.org/abs/1508.03624} {\path{arXiv:1508.03624}},
  \href {http://dx.doi.org/10.1007/JHEP02(2016)095}
  {\path{doi:10.1007/JHEP02(2016)095}}.

\bibitem{Bednyakov:2015}
A.~V. Bednyakov, A.~F. Pikelner, {Four-loop strong coupling $\beta$-function in
  the Standard Model. }\href {http://arxiv.org/abs/1508.02680}
  {\path{arXiv:1508.02680}}.

\bibitem{Chetyrkin:2016ruf}
K.~G. Chetyrkin, M.~F. Zoller, {Leading QCD-induced four-loop contributions to
  the $\beta$-function of the Higgs self-coupling in the SM and vacuum
  stability. }\href {http://arxiv.org/abs/1604.00853}
  {\path{arXiv:1604.00853}}.

\bibitem{Coleman:1973jx}
S.~R. Coleman, E.~J. Weinberg, {Radiative Corrections as the Origin of
  Spontaneous Symmetry Breaking}, Phys. Rev. D7 (1973) 1888--1910.
\newblock \href {http://dx.doi.org/10.1103/PhysRevD.7.1888}
  {\path{doi:10.1103/PhysRevD.7.1888}}.

\bibitem{Elias-Miro:2014pca}
J.~Elias-Miro, J.~R. Espinosa, T.~Konstandin, {Taming Infrared Divergences in
  the Effective Potential}, JHEP 08 (2014) 034.
\newblock \href {http://arxiv.org/abs/1406.2652} {\path{arXiv:1406.2652}},
  \href {http://dx.doi.org/10.1007/JHEP08(2014)034}
  {\path{doi:10.1007/JHEP08(2014)034}}.

\bibitem{Nielsen:1975fs}
N.~K. Nielsen, {On the Gauge Dependence of Spontaneous Symmetry Breaking in
  Gauge Theories}, Nucl. Phys. B101 (1975) 173.
\newblock \href {http://dx.doi.org/10.1016/0550-3213(75)90301-6}
  {\path{doi:10.1016/0550-3213(75)90301-6}}.

\bibitem{CMS:2014hta}
CMS, {Combination of the CMS top-quark mass measurements from Run 1 of the LHC
  (2014). Rep. number: CMS-PAS-TOP-14-015}.

\bibitem{Masinatop}
I.~Masina, A.~Notari, {Standard Model False Vacuum Inflation: Correlating the
  Tensor-to-Scalar Ratio to the Top Quark and Higgs Boson masses},
  Phys.Rev.Lett. 108 (2012) 191302.
\newblock \href {http://arxiv.org/abs/1112.5430} {\path{arXiv:1112.5430}},
  \href {http://dx.doi.org/10.1103/PhysRevLett.108.191302}
  {\path{doi:10.1103/PhysRevLett.108.191302}}.

\bibitem{Masinahybrid}
I.~Masina, A.~Notari, {Inflation from the Higgs field false vacuum with hybrid
  potential}, JCAP 1211 (2012) 031.
\newblock \href {http://arxiv.org/abs/1204.4155} {\path{arXiv:1204.4155}},
  \href {http://dx.doi.org/10.1088/1475-7516/2012/11/031}
  {\path{doi:10.1088/1475-7516/2012/11/031}}.

\bibitem{Masinaupgrade}
I.~Masina, {The Gravitational Wave Background and Higgs False Vacuum
  Inflation}, Phys.Rev. D89 (2014) 123505.
\newblock \href {http://arxiv.org/abs/1403.5244} {\path{arXiv:1403.5244}},
  \href {http://dx.doi.org/10.1103/PhysRevD.89.123505}
  {\path{doi:10.1103/PhysRevD.89.123505}}.

\bibitem{Ade:2015xua}
P.~A.~R. Ade, et~al., {Planck 2015 results. XIII. Cosmological parameters.
  }\href {http://arxiv.org/abs/1502.01589} {\path{arXiv:1502.01589}}.

\bibitem{Ade:2015lrj}
P.~A.~R. Ade, et~al., {Planck 2015 results. XX. Constraints on inflation.
  }\href {http://arxiv.org/abs/1502.02114} {\path{arXiv:1502.02114}}.

\bibitem{Ade:2015tva}
P.~Ade, et~al., {Joint Analysis of BICEP2/$Keck ?Array$ and $Planck$ Data},
  Phys. Rev. Lett. 114 (2015) 101301.
\newblock \href {http://arxiv.org/abs/1502.00612} {\path{arXiv:1502.00612}},
  \href {http://dx.doi.org/10.1103/PhysRevLett.114.101301}
  {\path{doi:10.1103/PhysRevLett.114.101301}}.

\bibitem{Bicep}
P.~Ade, et~al., {Detection of B-Mode Polarization at Degree Angular Scales by
  BICEP2}, Phys.Rev.Lett. 112 (2014) 241--101.
\newblock \href {http://arxiv.org/abs/1403.3985} {\path{arXiv:1403.3985}},
  \href {http://dx.doi.org/10.1103/PhysRevLett.112.241101}
  {\path{doi:10.1103/PhysRevLett.112.241101}}.

\bibitem{Branchina:2013jra}
V.~Branchina, E.~Messina, {Stability, Higgs Boson Mass and New Physics}, Phys.
  Rev. Lett. 111 (2013) 241801.
\newblock \href {http://arxiv.org/abs/1307.5193} {\path{arXiv:1307.5193}},
  \href {http://dx.doi.org/10.1103/PhysRevLett.111.241801}
  {\path{doi:10.1103/PhysRevLett.111.241801}}.

\bibitem{Branchina:2014usa}
V.~Branchina, E.~Messina, A.~Platania, {Top mass determination, Higgs
  inflation, and vacuum stability}, JHEP 09 (2014) 182.
\newblock \href {http://arxiv.org/abs/1407.4112} {\path{arXiv:1407.4112}},
  \href {http://dx.doi.org/10.1007/JHEP09(2014)182}
  {\path{doi:10.1007/JHEP09(2014)182}}.

\bibitem{Branchina:2015nda}
V.~Branchina, E.~Messina, {Stability and UV completion of the Standard Model.
  }\href {http://arxiv.org/abs/1507.08812} {\path{arXiv:1507.08812}}.

\bibitem{Loebbert:2015eea}
F.~Loebbert, J.~Plefka, {Quantum Gravitational Contributions to the Standard
  Model Effective Potential and Vacuum Stability}, Mod. Phys. Lett. A30~(34)
  (2015) 1550189.
\newblock \href {http://arxiv.org/abs/1502.03093} {\path{arXiv:1502.03093}},
  \href {http://dx.doi.org/10.1142/S0217732315501898}
  {\path{doi:10.1142/S0217732315501898}}.

\bibitem{Lalak:2015bwa}
Z.~Lalak, M.~Lewicki, P.~Olszewski, {Standard Model vacuum stability in the
  presence of gauge invariant nonrenormalizable operators}, in: {18th
  International Conference From the Planck Scale to the Electroweak Scale
  (Planck 2015) Ioannina, Greece, May 25-29, 2015}, 2015.
\newblock \href {http://arxiv.org/abs/1510.02797} {\path{arXiv:1510.02797}}.

\bibitem{Espinosa:2015kwx}
J.~R. Espinosa, {Implications of the top (and Higgs) mass for vacuum
  stability}, in: {8th International Workshop on Top Quark Physics (TOP2015)
  Ischia, NA, Italy, September 14-18, 2015}, 2015.
\newblock \href {http://arxiv.org/abs/1512.01222} {\path{arXiv:1512.01222}}.

\bibitem{EliasMiro:2012ay}
J.~Elias-Miro, J.~R. Espinosa, G.~F. Giudice, H.~M. Lee, A.~Strumia,
  {Stabilization of the Electroweak Vacuum by a Scalar Threshold Effect}, JHEP
  06 (2012) 031.
\newblock \href {http://arxiv.org/abs/1203.0237} {\path{arXiv:1203.0237}},
  \href {http://dx.doi.org/10.1007/JHEP06(2012)031}
  {\path{doi:10.1007/JHEP06(2012)031}}.

\end{thebibliography}
\end{document}